\documentclass[a4paper,fleqn]{cas-sc}

\usepackage[authoryear]{natbib}

\usepackage{booktabs}
\usepackage{multirow}
\usepackage[table]{xcolor}

\usepackage{tabularx}
\usepackage{array}
\usepackage{amsmath}
\usepackage{makecell}

\usepackage{amssymb}
\usepackage{arydshln}

\usepackage{xcolor}
\def\tsc#1{\csdef{#1}{\textsc{\lowercase{#1}}\xspace}}
\tsc{WGM}
\tsc{QE}

\begin{document}
\let\WriteBookmarks\relax
\def\floatpagepagefraction{1}
\def\textpagefraction{.001}

\shorttitle{}    

\shortauthors{}  

\title [mode = title]{From Continuous sEMG Signals to Discrete Muscle State Tokens: A Robust and Interpretable Representation Framework}  


\author[1,2]{Yuepeng Chen}
\fnmark[1]




\credit{Conceptualization, Methodology, Writing – original draft}

\affiliation[1]{organization={School of Computer Science (National Pilot Software Engineering School), Beijing University of Posts and Telecommunications},
            city={Beijing},
            country={China}}

\affiliation[2]{organization={Key Laboratory of Trustworthy Distributed Computing and Service (BUPT), Ministry of Education},
            city={Beijing},
            country={China}}

\affiliation[3]{organization={Department of Electronic Engineering, Tsinghua University},
            city={Beijing},
            country={China}}

\affiliation[4]{organization={Institute for Precision Medicine, Tsinghua University},
            city={Beijing},
            country={China}}

\affiliation[5]{organization={College of AI, Tsinghua University},
            city={Beijing},
            country={China}}

\affiliation[6]{organization={Research Academy of Grand Health, Ningbo University},
            city={Ningbo},
            country={China}}

\affiliation[7]{organization={School of Information Technology and Artificial Intelligence, Zhejiang University of Finance and Economics},
            city={Hangzhou},
            country={China}}
            
\author[3]{Kaili Zheng}
\fnmark[1]
\credit{Writing–review\&editing, Validation,Methodology}

\author[3,4,5]{Ji Wu}[orcid=0000-0001-6170-726X]
\ead{wuji_ee@tsinghua.edu.cn}

\cormark[1]
\credit{Supervision, Writing–review\&editing}

\author[3]{Zhuangzhuang Li}
\credit{Data curation, Investigation}

\author[6]{Ye Ma}
\credit{Methodology, Formal analysis}

\author[7]{Dongwei Liu}
\credit{Methodology, Formal analysis}

\author[3,4]{Chenyi Guo}[orcid=0000-0002-8222-5179]
\ead{guochy@mail.tsinghua.edu.cn}
\cormark[1]
\credit{Supervision, Writing–review\&editing}

\author[1,2]{Xiangling Fu}[orcid=0000-0002-1492-2829]
\ead{fuxiangling@bupt.edu.cn}
\cormark[1]
\credit{Supervision, Writing–review\&editing}

\cortext[1]{Corresponding author}
\fntext[1]{These authors contributed equally to this work.}

\begin{abstract}
Surface electromyography (sEMG) provides a non-invasive measure of neuromuscular activity, yet its continuous waveforms exhibit substantial inter-subject variability, making muscle activities difficult to compare across individuals. Although existing methods improve cross-subject action recognition, they mainly learn action-category invariance and remain limited in characterizing the correspondence of muscle-activity changes across individuals. Motivated by this gap, we propose a Muscle-State Tokenization Framework that constructs a cross-subject shared muscle-state vocabulary and converts continuous sEMG signals into discrete token sequences. We also introduce a large-scale dataset, ActionEMG-43, comprising 43 diverse actions and sEMG recordings from 16 major muscle groups across the body. Experiments on ActionEMG-43 demonstrate that the proposed tokens achieve high inter-subject consistency (Cohen’s Kappa = 0.82 $\pm$ 0.09), provide interpretable muscle-state representations, and outperform sEMG signals across all evaluated models in cross-subject action recognition.
These results suggest that muscle-state tokenization can mitigate the impact of inter-subject variability while preserving task-relevant information, providing a robust and interpretable representation layer for sEMG analysis with potential applications in rehabilitation assessment, human–machine interaction, and motor function analysis.

\end{abstract}




\maketitle

\section{Introduction}\label{sec1}
Surface electromyography (sEMG) provides a non-invasive way to capture neuromuscular activity from the skin surface and has been widely used in human action recognition, human--machine interaction, rehabilitation assessment, and wearable motion monitoring \citep{9751758,10254204}. Unlike vision-based or motion-capture systems that mainly describe external body kinematics, sEMG reflects the underlying muscle activation patterns associated with movement generation. This physiological property makes sEMG particularly valuable for analyzing how actions are produced, rather than only observing what actions are performed. With the development of wearable sensing technologies, multi-channel sEMG has become increasingly feasible for monitoring muscle activity during daily and whole-body movements, offering new opportunities for robust and interpretable human movement analysis \citep{AlAyyad2023,Wang2023}. 

Despite its potential, sEMG-based movement understanding remains challenging. Continuous sEMG signals are highly variable across individuals due to differences in muscle anatomy, subcutaneous tissue, motor control strategies, skin-electrode contact, and habitual movement patterns \citep{WANG2024106086}. Even when different individuals perform the same action, the recorded sEMG signals may exhibit considerable differences in amplitude, temporal dynamics, and spectral characteristics \citep{AITYOUS2025109651,ZHANG2026114517}. In addition, sEMG signals are sensitive to acquisition conditions such as electrode placement, sensor configuration, and muscle fatigue \citep{AO2024122304,10.3389/fnins.2025.1750792}. These factors make it difficult to learn representations that are both robust to individual differences and interpretable in terms of muscle activity.

Previous studies have attempted to improve the cross-subject robustness of sEMG from different perspectives. At the signal preprocessing level, MVC normalization is one of the most commonly used calibration strategies\citep{martens2015intra,diong2022muscle}. This method scales the amplitude of sEMG signals according to each individual’s maximal voluntary contraction (MVC) level, thereby reducing amplitude-scale differences across individuals and improving signal comparability. However, MVC normalization relies on participants accurately performing maximal voluntary contraction tests. Therefore, its reliability may be compromised in rehabilitation populations, individuals with reduced muscle strength, or data collection settings without professional supervision \citep{ZELLERS2019104}. At the model level, conventional methods typically extract handcrafted features from sEMG signals, such as root mean square, mean absolute value, and frequency-domain features, and combine them with classical classifiers such as SVM and KNN for action recognition \citep{cai2019svm,ZHOU2021102577}. These features can describe the amplitude, temporal morphology, and spectral characteristics of local muscle activity in a low-dimensional form and provide a certain degree of interpretability \citep{10254204}. In recent years, deep learning models, including convolutional neural networks, recurrent neural networks, and Transformers, have been introduced into sEMG-based action recognition. These models learn task-related representations from raw or preprocessed sEMG signals in an end-to-end manner \citep{hu2018novel,cai2021hybrid,montazerin2022vit,chen2025waveformer}. In addition, transfer learning \citep{cote2017transfer,lin2024reducing}, domain adaptation \citep{du2017surface,9854078,wang2024optimization}, and contrastive learning \citep{yang2025stcnet} methods have also been used to mitigate cross-subject distribution differences. These methods typically learn representations with stronger cross-subject robustness through feature alignment, target-domain adaptation, or label supervision, thereby improving recognition performance on unseen individuals to some extent.

\begin{figure}
    \centering
    \includegraphics[width=.65\textwidth]{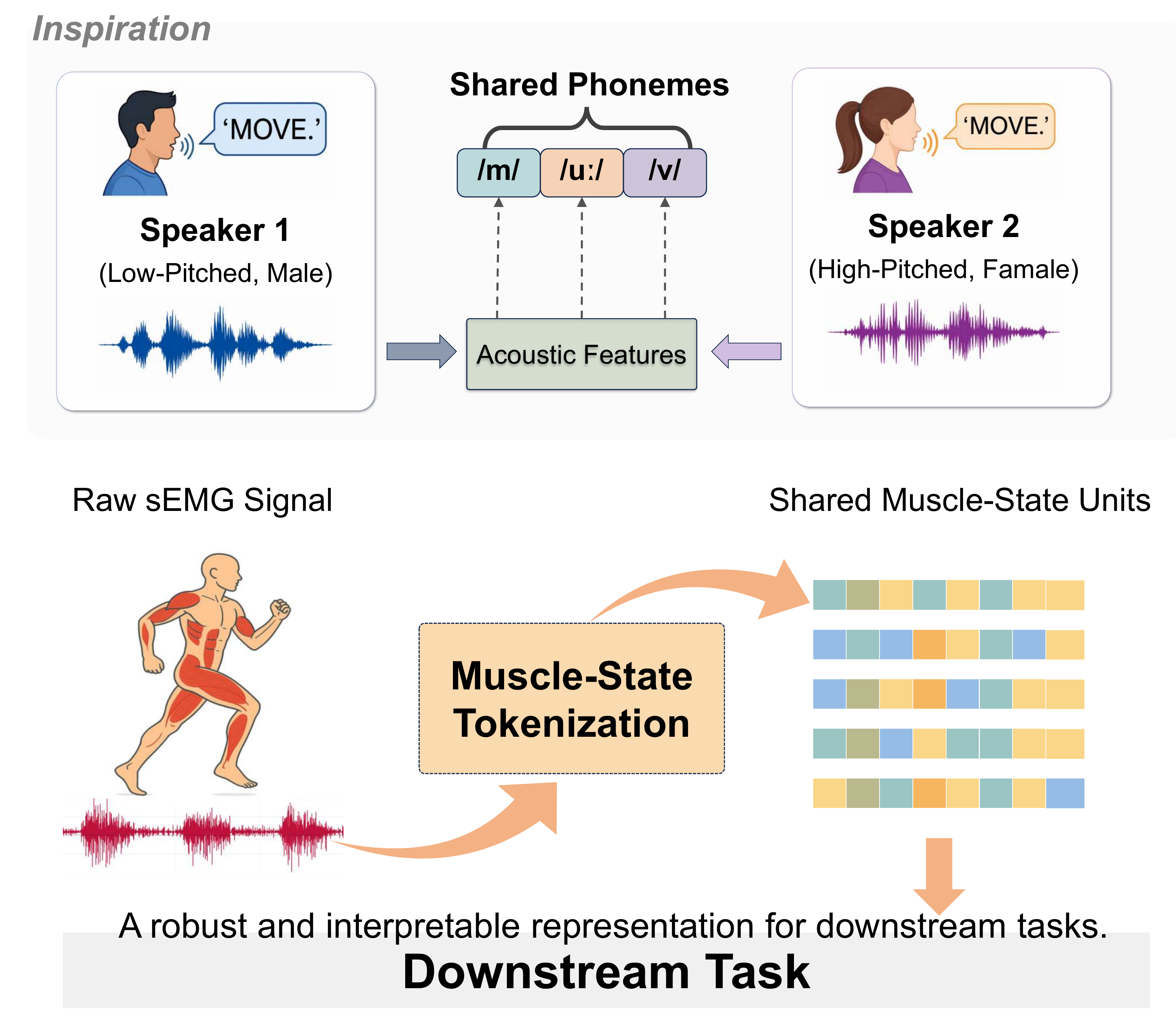}
    \caption{Overview of the proposed sEMG tokenization framework. Inspired by speech recognition, where variable acoustic waveforms can be mapped to shared phoneme-like units, we propose to transform continuous sEMG signals into discrete muscle-state units.}
    \label{fig:overall-method}
\end{figure}

Although the above methods can improve the robustness of sEMG signals on unseen individuals, the learned cross-subject invariance is mainly reflected in discriminative consistency at the action-category level, making it difficult to explicitly characterize the correspondence of muscle-activity changes across individuals performing the same action. This limitation is important because applications such as rehabilitation assessment, motor function analysis, and movement quality monitoring require not only recognizing the performed action, but also describing the underlying muscle activity in a way that can be consistently interpreted across different users, sessions, and execution conditions. Based on this observation, this study further explores, from the perspective of cross-subject comparability of muscle activities, whether sEMG signals contain muscle-state units that can be shared across individuals, thereby providing a new perspective for mitigating the impact of individual differences in sEMG signals. This idea is inspired by speech processing, where continuous acoustic signals vary substantially across speakers in pitch, accent, speaking rate, and speaking style, yet can be described using relatively stable discrete units, such as phoneme-like representations. Similarly, despite substantial inter-subject differences in sEMG signals, short-time muscle activities may exhibit shared latent structures characterized by activation intensity, waveform complexity, and spectral composition. sEMG signals can be transformed into a representation that is more comparable across individuals and more interpretable in terms of muscle activity. The conceptual motivation of the proposed framework is illustrated in Figure~\ref{fig:overall-method}.


Based on this motivation, we propose an sEMG tokenization framework that converts continuous sEMG signals into sequences of discrete muscle-state tokens. Specifically, each sEMG signal is segmented using short sliding windows, and multiple time-domain, frequency-domain, and autoregressive features are extracted from each local window to characterize the instantaneous muscle activity. A data-driven clustering procedure is then used to discover a vocabulary of muscle-state units in the feature space. Each local sEMG window is assigned to its nearest muscle-state unit, thereby converting continuous sEMG signals into discrete token sequences. To evaluate the proposed representation, we constructed ActionEMG-43, a multi-channel sEMG dataset containing 43 whole-body actions performed by 14 healthy adults. Based on this dataset, we evaluated sEMG tokens in terms of inter-subject consistency, interpretability, and action-recognition performance. The results show that the learned tokens achieved stable cross-subject assignment consistency (Cohen’s Kappa = 0.82 ± 0.09) and exhibited distinct and interpretable sEMG feature profiles. In particular, in the cross-subject action-recognition task, we evaluated a range of conventional machine-learning and deep-learning models. The sEMG tokens consistently achieved significantly better performance than raw sEMG signals across different models, suggesting that the proposed token representation can preserve task-relevant information while mitigating the influence of substantial inter-subject variability. These findings suggest that sEMG tokens provide a robust and interpretable representation of muscle activity.

The main contributions of this work are summarized as follows:

\begin{itemize}

\item We present a new perspective for sEMG signals analysis by exploring whether continuous sEMG signals can be organized into shared muscle-state units across different individuals.

\item We propose an sEMG tokenization framework that discretizes continuous sEMG signals into muscle-state tokens.

\item We introduce ActionEMG-43, a multi-channel sEMG dataset covering 43 whole-body actions and 16 muscle channels across the upper and lower limbs. 

\item Experimental results demonstrate that sEMG tokens provide a cross-subject robust, interpretable, and action-discriminative representation for sEMG-based movement analysis.

\end{itemize}

\section{Related Work}\label{sec1-1}

\subsection{Handcrafted Feature Extraction for sEMG signals}

Handcrafted feature extraction has long been a key component of sEMG signal analysis. Since raw sEMG signals are stochastic and non-stationary, and are easily affected by electrode placement, skin contact conditions, muscle fatigue, and inter-subject variability, directly analyzing raw waveforms often makes it difficult to obtain stable results \citep{tkach2010study,phinyomark2018feature}. Therefore, existing studies typically segment continuous sEMG signals into short analysis windows and extract handcrafted features\citep{zheng2022review}. Handcrafted sEMG features can generally be categorized into time-domain, frequency-domain, time-frequency-domain, and model-based features. Time-domain features are the most widely used, such as root mean square, mean absolute value, waveform length, zero crossing, slope sign change, and Willison amplitude, were widely used to describe muscle activation intensity and waveform complexity, whereas frequency-domain features, such as mean frequency and median frequency, were adopted to characterize spectral changes associated with muscle activation and fatigue \citep{PHINYOMARK20127420,muceli2024tutorial}.

The extracted handcrafted features are mainly used for two purposes. First, in movement analysis and rehabilitation assessment, sEMG features serve as quantitative indicators for characterizing muscle activation patterns, fatigue-related changes, neuromuscular function, and motor-control differences. For example, amplitude-related and spectral features have been used to analyze muscle recruitment, abnormal activation, compensatory movement, and rehabilitation progress during dynamic tasks such as gait and exercise \citep{campanini2020surface,agostini2020surface,sun2022application}. Second, in recognition-oriented studies, handcrafted features are commonly used as compact and discriminative input representations for machine-learning classifiers. By reducing high-dimensional raw waveforms into low-dimensional feature vectors, these features can improve computational efficiency and recognition performance in hand gesture recognition, prosthetic control, human--machine interaction, and action-recognition tasks \citep{jaramillo2020real,sultana2023systematic}.

\subsection{Representation Learning for sEMG signals}
Deep learning methods have been widely introduced to learn task-related representations directly from raw or preprocessed sEMG signals. Hu et al. proposed an attention-based hybrid CNN-RNN architecture for sEMG-based recognition task, where convolutional layers extract local representations and recurrent layers capture temporal dynamics \citep{hu2018novel}. Similarly, CNN-LSTM architectures combine local feature extraction with long-term temporal modeling for hand gesture recognition \citep{cai2021hybrid}. More recently, Transformer-based methods have been adopted to model global dependencies among channels and temporal segments. ViT-HGR applies a Vision Transformer structure to high-density sEMG signals, treating the spatial arrangement of electrodes as structured input patches \citep{montazerin2022vit}. WaveFormer further introduces a lightweight Transformer framework for sEMG-based recognition task, aiming to improve efficiency while preserving time-frequency representation capacity \citep{chen2025waveformer}. These methods demonstrate the ability of deep models to learn expressive continuous embeddings from sEMG data. Transfer learning and domain adaptation have also been explored to mitigate cross-subject distribution discrepancies. Côté-Allard et al. introduced transfer learning for sEMG recognition task using CNNs, showing that knowledge learned from multiple users can improve the adaptation of deep models to new individuals \citep{cote2017transfer}. Du et al. used deep domain adaptation to enhance inter-session sEMG action recognition by reducing distribution shifts between different recording sessions \citep{du2017surface}. Shi et al. improved the robustness and adaptability of sEMG-based pattern recognition through deep domain adaptation, and Wang et al. further investigated unsupervised domain adaptation for inter-subject action recognition \citep{9854078,wang2024optimization}. These methods improve generalization by learning domain-invariant feature spaces, often through discrepancy minimization, adversarial alignment, or target-domain adaptation. However, many of them still require target-domain data during adaptation. Contrastive learning provides another way to improve cross-subject robustness by constructing positive and negative pairs in the representation space. In sEMG recognition, samples from the same action category across different subjects can be pulled closer, while samples from different categories are pushed apart. For example, STCNet introduces a spatio-temporal cross network with subject-aware contrastive learning to improve action recognition by encouraging subject-robust action representations \citep{yang2025stcnet}. Such methods can reduce subject-specific variations and improve recognition performance on unseen individuals. 

Although existing cross-subject representation learning methods have improved sEMG-based action recognition, they mainly focus on aligning samples with the same action label across individuals. As a result, the learned representations tend to capture action-level invariance rather than the detailed muscle-activity process underlying action execution. This may obscure fine-grained inter-subject differences in local muscle activation patterns and limits the ability to compare how different individuals perform the same movement.



\section{Material and Methods}\label{sec2}
\subsection{Dataset}
\subsubsection{ActionEMG-43 Dataset}
\label{sec:results-actionemg43}
\begin{figure}
    \centering
    \includegraphics[width=0.9\linewidth]{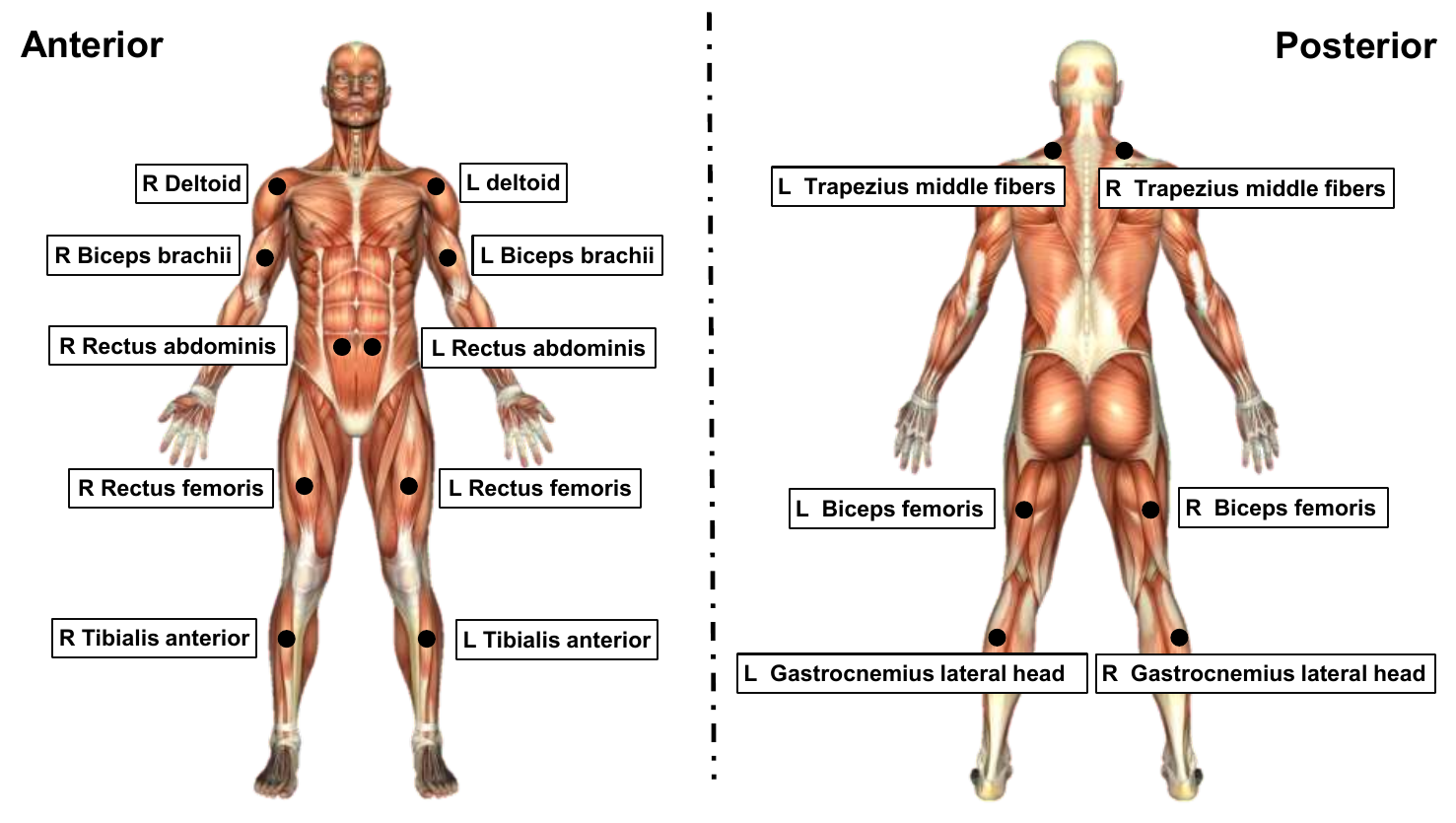}
    \caption{Muscles for sEMG Signals Collection in ActionEMG-43. sEMG signals were recorded from 16 major muscle groups across the upper limbs, lower limbs, and trunk using Delsys Trigno wireless sensors. This comprehensive coverage enables detailed analysis of whole-body motor coordination and muscle states across a wide range of actions.}
    \label{bodymuscle}
\end{figure}

To support research in sEMG-based motion analysis, we present ActionEMG-43, a large-scale dataset designed to benchmark a wide range of tasks including action recognition, tokenization-based representation learning, and movement quality assessment. As presented in Table~\ref{tab:dataset_comparison}, ActionEMG-43 is the most comprehensive and versatile sEMG dataset to date. It comprises 43 distinct actions, encompassing a broad spectrum of daily and exercise-related movements such as jogging, high knee lifts, frog jumps, high kicks, and squats. A full list of the actions is provided in Appendix Table.~\ref{ActionEMG-43}. The data were collected from 14 healthy adult participants (aged 21–36 years, mean ± std: 27.7 ± 5.1 years), with each subject performing at least three repetitions of each cyclical movement and maintaining each static or continuous movement for a minimum of two seconds. The sEMG signals were recorded using Delsys Trigno wireless sensors from 16 major muscle groups, providing extensive coverage of both upper and lower limbs as well as trunk muscles. Sensor placement is illustrated in Fig.~\ref{bodymuscle}. The signals were sampled at 1259 Hz, ensuring high temporal resolution for detailed time-frequency analysis.

\subsubsection{MEQAD Dataset}
To explore whether the sEMG tokens can support interpretable analysis of movement execution, we additionally collected a focused Movement Quality Assessment Dataset (MEQAD). Whereas ActionEMG-43 was developed for cross-subject representation learning and action recognition, MEQAD serves as a small-scale case-study dataset for examining fine-grained differences between standard and faulty movement executions. MEQAD contains three representative actions: High Knees, Squats, and Biceps Curls. For each action, both a correct execution and a commonly observed faulty execution were defined in advance. One participant, a Certified Strength and Conditioning Specialist accredited by the National Strength and Conditioning Association, performed both versions as reference demonstrations. Two participants without professional training backgrounds subsequently performed the same movement variants under expert instruction and supervision. The same sEMG system, sensor placement, muscle coverage, and sampling rate as those used for ActionEMG-43 were adopted. Detailed definitions of the correct and faulty movement variants and the corresponding instructions are provided in Appendix~A.

\begin{table}[ht!]
\caption{Comparison of publicly available sEMG datasets in terms of subject count, anatomical coverage, number of muscle groups, and action diversity. ActionEMG-43 offers the most comprehensive coverage, spanning the whole body with 16 muscle groups and 43 distinct actions, providing a benchmark for large-scale motor analysis.}
    \label{tab:dataset_comparison}
    \centering
    \begin{tabular}{lc|cccc}
    \toprule
        Name & Year & \# Subjects & Body Region & \# Muscle Groups & \# Actions \\ \midrule
        EMAHA-DB1 \citep{EMAHA-DB1} & 2023 & 24 & right arm & 5 & 22 \\
        MyoBit \citep{MyoBit} & 2023 & 7 & elbow & 1 & 7 \\
        SIAT-LLMD \citep{wei2023surface} & 2023 & 40 & left leg & 9 & 16 \\
        JJ dataset \citep{JJDataset} & 2024 & 15 & left leg & 9 & 10 \\
        ULTRA-MoCap \citep{fritsche2025ultra} & 2025 & 13 & upper body & 3 & 5 \\
        \textbf{ActionEMG-43} & 2025 & 14 & \textbf{whole body} & \textbf{16} & \textbf{43} \\
    \bottomrule
    \end{tabular}
    
\end{table}

\subsubsection{Dataset Split}

Based on the two datasets described above, we systematically evaluate the proposed tokenization framework across multiple dimensions. Using ActionEMG-43, we assess the inter-subject consistency and representation capacity of sEMG tokens in large-scale action recognition tasks. In addition, MEQAD enables fine-grained analysis of movement quality, allowing us to validate the interpretability of tokenized representations in capturing underactivation, compensatory patterns, and other execution-level differences. In our experiments, we employ a five-fold cross-validation strategy unless otherwise specified. In each fold, samples from a subset of subjects are used for validation, and samples from the remaining subjects were used for training. Both the cluster centers and the downstream-task models are trained on the training set, and the experimental results reported in this paper are based on the validation set. This approach ensures that the training and validation subsets are completely separate, preventing any information leakage during both the tokenization and model training processes.

\subsection{Muscle-State Tokenization}\label{sec:method-tokenization}
\begin{figure}
    \centering
    \includegraphics[width=0.9\linewidth]{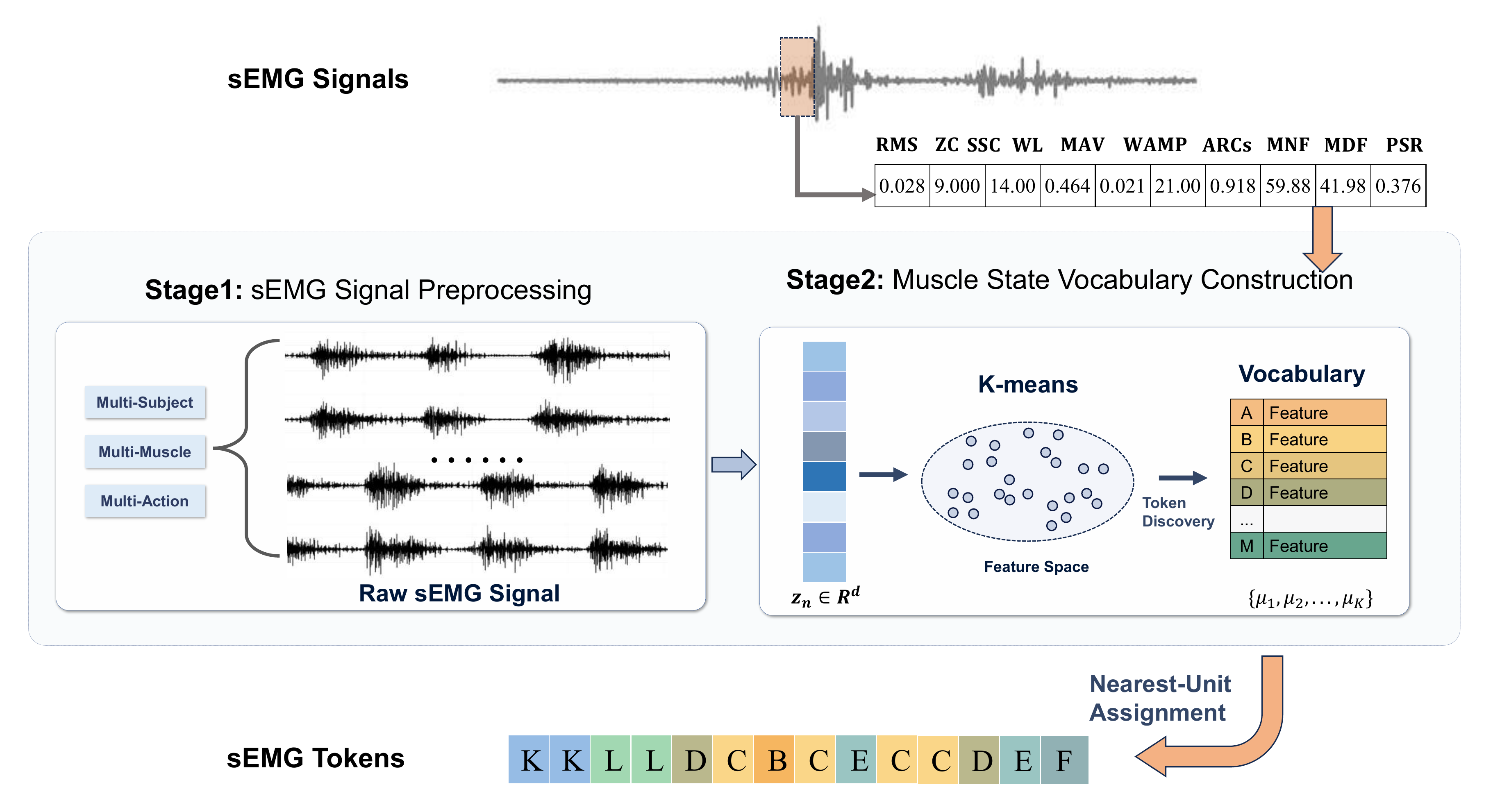}
    \caption{Muscle-state tokenization framework}
    \label{Figure_12}
\end{figure}

We propose a muscle-state tokenization framework that transforms continuous multi-channel sEMG signals into discrete sEMG tokens. As illustrated in Fig.~\ref{Figure_12}, the framework comprises three stages: sEMG preprocessing, Muscle State Vocabulary Construction, and discrete token assignment. 

\subsubsection{sEMG Signal Preprocessing}

Given a multi-channel sEMG recording, our objective is to convert the continuous and subject-dependent signal into a sequence of discrete sEMG token. To identify fine-grained muscle-state units that recur consistently within continuous sEMG signals, we first decompose each muscle channel into a sequence of short-term windows and represent each window using physiologically interpretable descriptors. The raw sEMG signals are first processed using a fourth-order Butterworth band-pass filter with cutoff frequencies of 20 and 450 Hz. This preprocessing suppresses low-frequency motion artifacts and high-frequency interference while preserving the frequency components primarily associated with muscle electrical activity, following common practice in sEMG analysis \citep{moreira2021lower,lu2022continuous,lim2024assessment,11000291}. The filtered signals are subsequently divided into overlapping short-term windows. A fundamental mechanical response of muscle fibers to brief stimulation is a muscle twitch, which typically lasts approximately 100 ms and consists of a latent period, a contraction phase, and a relaxation phase \citep{jenkins2016anatomy,carlson2018human}. Although the duration of an isolated muscle twitch cannot be directly equated with the appropriate analysis window for sEMG, this timescale provides a useful physiological reference for selecting a short local analysis window. A 50-ms window, shorter than the approximate twitch timescale, allows local changes in muscle activation to be characterized. We therefore used a 50-ms sliding window with a 25-ms stride, and further validated this choice through the window-length sensitivity analysis reported in Section~\ref{window-length}.

\subsubsection{Muscle State Vocabulary Construction}
Each window is represented using ten window-level descriptors with physiologically interpretable: root mean square (RMS), zero crossings (ZC), slope sign changes (SSC), waveform length (WL), mean absolute value (MAV), Willison amplitude (WAMP), autoregressive coefficient (ARC), mean frequency (MNF), median frequency (MDF), and power spectrum ratio (PSR). These descriptors have been widely used in sEMG signal analysis \citep{duan2023hybrid,atzori2014electromyography}. Specifically, RMS and MAV characterize the overall signal amplitude and are commonly associated with the local level of muscle activation. WL quantifies cumulative waveform variation and is influenced by both signal magnitude and temporal fluctuation. ZC, SSC, and WAMP characterize waveform oscillation, local slope reversals, and threshold-qualified changes between adjacent samples, respectively. ARC describes short-term temporal dependence within the local waveform. MNF, MDF, and PSR characterize the distribution and concentration of spectral energy and may capture changes associated with motor-unit activity and fatigue-related spectral shifts. Together, these descriptors define a physiology-interpretable state space that jointly represents activation amplitude, waveform complexity, short-term temporal dependence, and spectral organization. Their mathematical definitions are provided in Table~\ref{tab:semg_feature_definitions}.

Let $\mathbf{z}_n = \phi(\mathbf{x}_n) \in \mathbb{R}^{d}$ denote the descriptor vector extracted from the $n$-th local sEMG window, where $\phi(\cdot)$ represents the descriptor-extraction operation and $d$ is the descriptor dimension. Before vocabulary construction, each descriptor dimension is standardized using the mean and standard deviation estimated exclusively from the training set. Although the resulting descriptor space provides an interpretable basis for comparing local muscle activity, it remains continuous and contains a large number of closely related window-level observations. We therefore summarize the standardized descriptors using a finite set of reusable muscle-state units. Let
\begin{equation} \widetilde{\mathcal{Z}} = \left\{ \widetilde{\mathbf{z}}_1, \widetilde{\mathbf{z}}_2, \ldots, \widetilde{\mathbf{z}}_{N_{\mathrm{tr}}} \right\}, \qquad \widetilde{\mathbf{z}}_n\in\mathbb{R}^{d}, \end{equation}
denote the standardized descriptor vectors extracted from all local windows in the training set, where $N_{\mathrm{tr}}$ is the total number of training windows.

K-means is applied to partition these descriptors into $K$ compact groups by minimizing the within-cluster variance: \begin{equation} \left\{ C_k^{*}, \boldsymbol{\mu}_k^{*} \right\}_{k=1}^{K} = \underset{ \{C_k,\boldsymbol{\mu}_k\}_{k=1}^{K} }{ \arg\min } \sum_{k=1}^{K} \sum_{\widetilde{\mathbf{z}}_n\in C_k} \left\| \widetilde{\mathbf{z}}_n-\boldsymbol{\mu}_k \right\|_2^2, \label{eq:muscle_state_vocabulary} \end{equation} where $C_k$ denotes the set of descriptors assigned to the $k$-th cluster and $\boldsymbol{\mu}_k\in\mathbb{R}^{d}$ is the corresponding centroid.

Since clustering is performed in the interpretable descriptor space, each centroid summarizes a characteristic combination of local amplitude, waveform, temporal, and spectral properties. We therefore treat $\boldsymbol{\mu}_k$ as the prototype of the $k$-th muscle-state unit. The complete prototype set \begin{equation} \mathcal{V} = \left\{ \boldsymbol{\mu}_1, \boldsymbol{\mu}_2, \ldots, \boldsymbol{\mu}_K \right\} \end{equation} constitutes the muscle-state vocabulary. 

The vocabulary is learned exclusively from the training set. Once constructed, both the prototypes and the feature-standardization statistics remain fixed during validation and inference. To facilitate the interpretation of relationships among the learned muscle-state units, the prototypes were further reordered according to their similarity in the descriptor space. Specifically, hierarchical clustering was performed on the prototype centroids using Euclidean distance. Because the leaf order of a hierarchical clustering tree is not unique, optimal leaf ordering (OLO) was subsequently applied to determine a unique ordering while preserving the hierarchical clustering structure. The resulting ordering was used only for visualization and analysis of token relationships, and did not affect vocabulary construction or token assignment.



\subsubsection{Token Assignment}

Given an unseen local sEMG window, the same preprocessing, feature-extraction, and standardization procedures are first applied to obtain its descriptor $\widetilde{\mathbf{z}}_t$. The window is then
assigned to the nearest muscle-state unit:
\begin{equation}
q_t
=
\arg\min_{k\in\{1,\ldots,K\}}
\left\|
\widetilde{\mathbf{z}}_t
-
\boldsymbol{\mu}_k
\right\|_2^2,
\label{eq:muscle_state_assignment}
\end{equation}
where $q_t$ is the discrete index of the muscle-state unit assigned to the $t$-th local window. Applying this operation to all $T$ windows of a recording yields the sEMG token sequence
\begin{equation}
\mathbf{q}
=
\left[
q_1,q_2,\ldots,q_T
\right].
\end{equation}

Through this procedure, continuously varying local sEMG descriptors are replaced by symbols drawn from a fixed vocabulary. Local windows with similar amplitude, waveform, temporal, and spectral profiles are mapped to the same muscle-state unit, while the temporal ordering of the original recording is retained in the resulting token sequence.







\subsection{Selection of the Number of Muscle-State Units}
The number of muscle-state units determines the granularity of the resulting discrete representation. A smaller number of muscle-state units $K$ produces a compact vocabulary but may partition the sEMG feature space too coarsely to preserve subtle variations in muscle activity. In contrast, a larger $K$ enables a finer-grained representation but may introduce subject-specific states, thereby reducing inter-subject consistency. We therefore select the appropriate value of $K$ from two complementary perspectives: inter-subject consistency and representation capacity.

\textbf{Inter-subject Consistency}
We evaluate consistency by comparing two sEMG token assignment strategies on the same test samples. In the first strategy, sEMG tokens are obtained by applying a clustering model trained on the training set. In the second, clustering is independently performed on the test set. The second strategy is used only for consistency analysis and is not involved in downstream model training or inference. 

Assume that the two sEMG token sequences are given by
\begin{equation}
\mathbf{y}^{(1)}
=
\left[
y_1^{(1)},\ldots,y_N^{(1)}
\right],
\qquad
\mathbf{y}^{(2)}
=
\left[
y_1^{(2)},\ldots,y_N^{(2)}
\right]
\end{equation}
where $N$ denotes the number of aligned evaluation windows, and $y_n^{(1)}$ and $y_n^{(2)}$ correspond to the same sEMG window under the two token assignment strategies.

Consistency is measured using Cohen's Kappa:
\begin{equation}
\kappa
=
\frac{p_o-p_e}{1-p_e}
\end{equation}
where \(p_o\) is the observed agreement, and \(p_e\) is the agreement
expected by chance based on the marginal distributions of the two label
sequences. Let \(\mathbf{O}\) denote the $K\times K$ contingency
matrix, where \(O_{i,j}\) is the number of samples labeled as category $i$ in the first sequence and category $j$ in the second sequence.

The observed agreement is computed as
\begin{equation}
p_o
=
\frac{1}{N}
\sum_{k=1}^{K}
O_{k,k}
\end{equation}

For the \(k\)-th category, the marginal proportions in the two
sequences are
\begin{equation}
p_k^{(1)}
=
\frac{O_{k,*}}{N},
\qquad
p_k^{(2)}
=
\frac{O_{*,k}}{N},
\qquad
\text{where}\quad
O_{k,*}
=
\sum_{j=1}^{K} O_{k,j},
\qquad
O_{*,k}
=
\sum_{i=1}^{K} O_{i,k}
\end{equation}

The agreement expected by chance is then computed as
\begin{equation}
p_e
=
\sum_{k=1}^{K}
p_k^{(1)}p_k^{(2)}
\end{equation}

A larger kappa indicates stronger agreement between the two token assignment strategies and, consequently, greater stability of the sEMG tokens. For descriptive interpretation, we refer to the conventional benchmarks proposed by Landis and Koch \citep{landis1977measurement}: values $\le$0 as no agreement, 0.01-0.20 as slight agreement, 0.21-0.40 as fair agreement, 0.41-0.60 as moderate agreement, 0.61-0.80 as substantial agreement, and 0.81-1.00 as almost perfect agreement.

\textbf{Representation Capacity}
we evaluate representation capacity of sEMG tokens using human action recognition task. Recognition performance is evaluated using accuracy, macro-averaged precision, macro-averaged recall, and macro-averaged F1 score.

Let $N_{\mathrm{sample}}$ denote the number of test samples and $N_{\mathrm{class}}$ denote the number of action classes. For the $i$-th sample, $y_i$ denotes its ground-truth label, and $\hat{y}_{i}$ denotes the predicted label. Accuracy is used to evaluate the proportion of correctly classified samples, as follows:

\begin{equation}
 Accuracy =\frac{\sum_{i=1}^{N_{sample}} \mathbf{1}\left[y_{i} = \hat{y}_{i}\right] }{N_{sample}}
\end{equation}

where $\mathbf{1}{\cdot}$ is the indicator function that evaluates to 1 if the predicted label is equal to the ground-truth label, and 0 otherwise.

For class $j$, precision $P_{j}$ is defined as the ratio of the number of query samples correctly predicted as class $j$ to the total number of query samples predicted as class $j$. The macro-averaged precision Macro-P, evaluates the model’s overall performance across all classes:
\begin{equation}
P_{j}=\frac{\sum_{i=1}^{N_{sample}} \mathbf{1}\left[\hat{y}_{i}=j \wedge y_{i}=j\right]}{\sum_{i=1}^{N_{sample}} \mathbf{1}\left[\hat{y}_{i}=j\right]}  , \qquad \text{Macro-P} = \frac{\sum_{j=1}^{N_{class}} P_{j}}{N_{class}} 
\end{equation}

For class $j$, recall $R_{j}$ measures the proportion of query samples correctly predicted as class $j$ against the total number of query samples that actually belong to class $j$. The macro-averaged recall Macro-R provides an overall performance indicator:
\begin{equation}
R_{j}=\frac{\sum_{i=1}^{N_{sample}} \mathbf{1}\left[\hat{y}_{i}=j \wedge y_{i}=j\right]}{\sum_{i=1}^{N_{sample}} \mathbf{1}\left[{y}_{i}=j\right]}  , \qquad \text{Macro-R} = \frac{\sum_{j=1}^{N_{class}} R_{j}}{N_{class}} 
\end{equation}

The F1 score combines precision and recall through their harmonic mean for each class $j$. The macro-averaged F1 score $\text{Macro-F1}$ averages the F1 scores across all classes:
\begin{equation}
F1_{j}=2 \times \frac{P_{j} \times R_{j}}{P_{j}+R_{j}}  , \qquad \text{Macro-F1} = \frac{\sum_{j=1}^{N_{class}} F1_{j}}{N_{class}} 
\end{equation}

All metrics range from 0 to 1, with higher values indicating better action classification performance.

\subsection{Downstream Task}

After determining the number of muscle-state units $K$, we further evaluate the applicability of the proposed sEMG token to discriminative and interpretive tasks through two downstream tasks: human action recognition and movement quality assessment.

\subsubsection{Human Action Recognition}
Unlike the comparative evaluation used during the selection of $K$, which examines the relative action-discriminative capacity of sEMG token representations generated with different numbers of muscle-state units, the present evaluation fixes the final value of $K$ and directly compares the resulting sEMG token with the raw sEMG signals. The aim is to determine whether the discrete representation retains sufficient action-discriminative information.

\begin{figure}
    \centering
    \includegraphics[width=\linewidth]{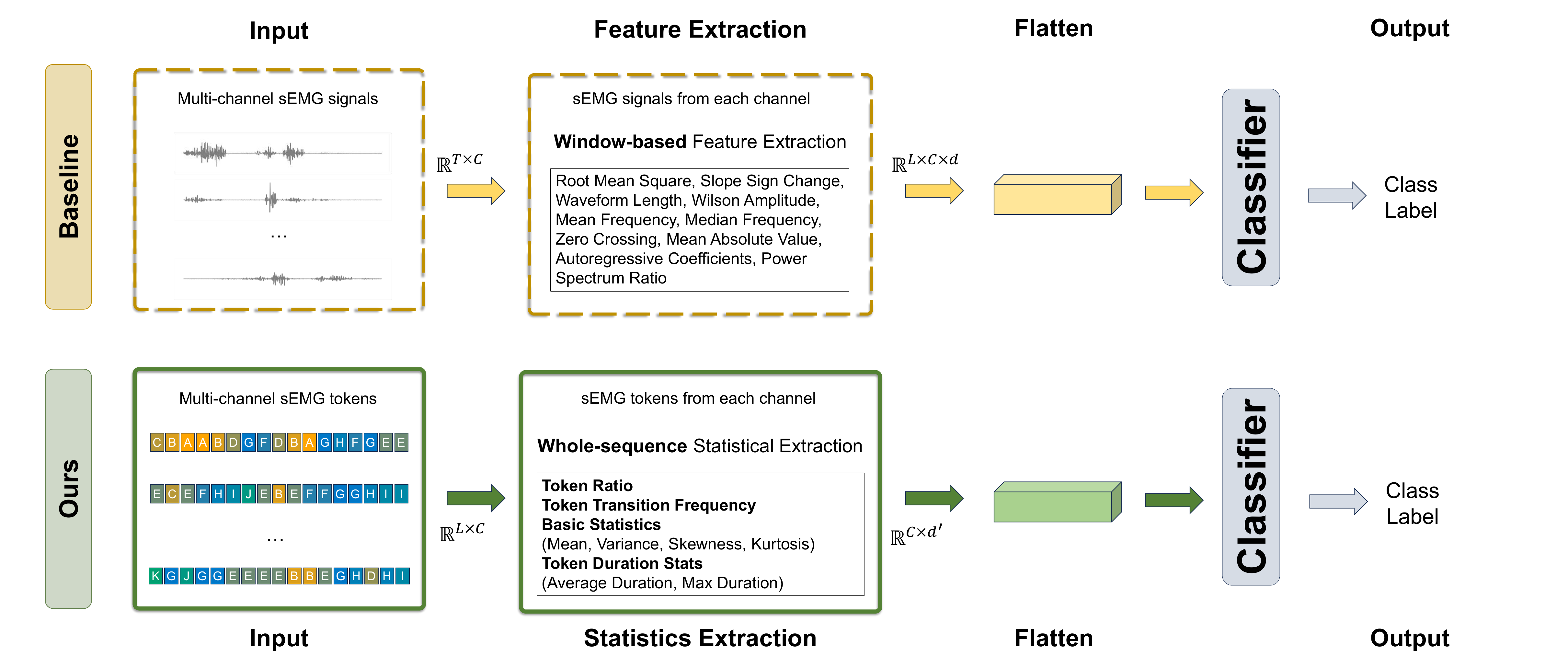}
    \caption{Conventional machine learning-based human action recognition framework. Handcrafted features extracted from raw sEMG signals and statistical descriptors derived from sEMG token sequences are separately flattened into fixed-length one-dimensional feature vectors and fed into an SVM or KNN classifier for action recognition.
    }
    \label{ActionRecognitionML}
\end{figure}

\begin{figure}
    \centering
    \includegraphics[width=\linewidth]{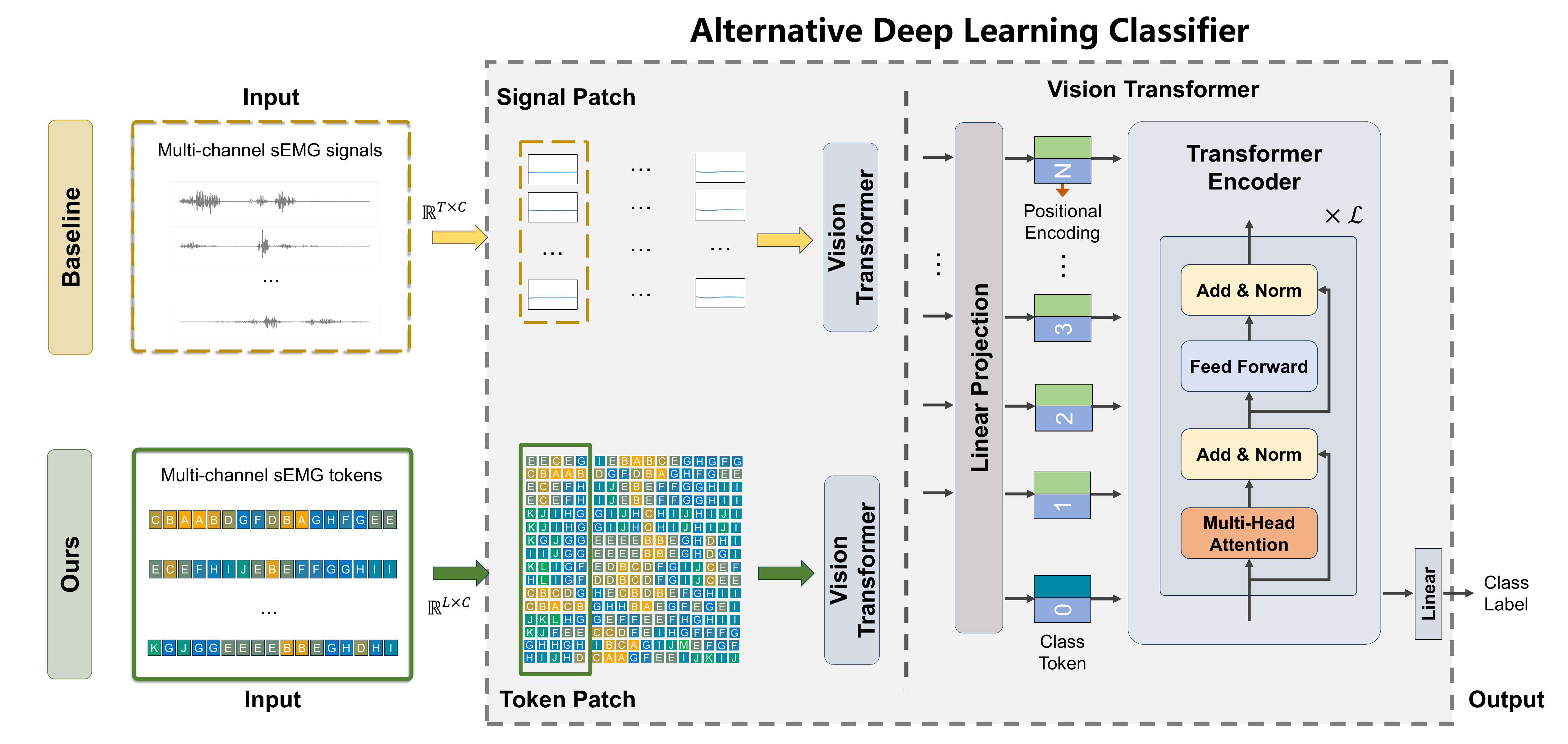}
    \caption{Deep learning-based human action recognition framework, with ViT-HGR shown as a representative implementation. For 16 muscles, the input sequence is either the raw sEMG signals or the corresponding sEMG tokens sequence. The sequence is divided into several temporal patches. Each patch is passed through a linear projection and positional embedding layer, and a class token is added to the beginning of the sequence. The resulting sequence is then processed by the transformer encoder. Raw sEMG signals are smoothed using a first-order low-pass Butterworth filter before segmentation.}
    \label{ActionRecognitionDL}
\end{figure}

We employ several representative models widely adopted in sEMG-based action recognition, covering both conventional feature-based classifiers and deep learning architectures. K-nearest neighbors (KNN) and support vector machine (SVM) are used as conventional feature-based classifiers, whereas ViT-HGR, WaveFormer, and STCNet are adopted as representative deep learning models. For KNN and SVM, handcrafted descriptors extracted from raw sEMG signals and statistical descriptors derived from sEMG token sequences are used as input features for classification. For deep learning architectures, the complete raw sEMG sequences and token sequences are directly processed by the networks to learn action-discriminative representations. Within each model, the data partition, model configuration, training protocol, and evaluation criteria are kept consistent between the signal-based and token-based settings. Figure~\ref{ActionRecognitionML} summarizes the feature construction and classification pipeline used by the conventional machine learning classifiers, whereas Figure~\ref{ActionRecognitionDL} presents the sequence representation learning and classification pipeline used by the deep learning models, with ViT-HGR shown as a representative example. As both models require fixed-length input, we standardize all input sequences to $L$ segments via replication-padding, which corresponds to padding raw sEMG signals to $T$ sampling points. Denote the number of muscle channels as $C$, as mentioned in Section~\ref{sec:results-actionemg43}. 

For the conventional machine learning classifiers, input representations are constructed as follows:
\begin{itemize}
    \item sEMG signals: $d$ commonly used time-frequency features (as described in Section~\ref{sec:method-tokenization}) are extracted from each segment of each muscle channel, forming a feature matrix of size $L \times C \times d$. 
    \item sEMG tokens: For each muscle, $d'$ statistical features are computed over the entire token sequence of length $L$, yielding a $C \times d'$ feature matrix. $d'$ statistical features include: token ratio (proportion of each token), token transition Frequency (frequency of token changes), token duration statistics (average and maximum continuous durations), and basic statistics (mean, variance, skewness, kurtosis).  And $mean =\frac{1}{L} \sum_{i=1}^{L} s_{i}; variance=\frac{1}{L} \sum_{i=1}^{L} (s_{i}-mean)^{2}; skewness =\frac{ \frac{1}{L} \sum_{i=1}^{L} (s_{i}-mean)^{3}}{(variance)^{3/2} }; kurtosis=\frac{ \frac{1}{L} \sum_{i=1}^{L} (s_{i}-mean)^{4}}{(variance)^{2} } -3$, where $s_i$ is the token value at time step $i$.
\end{itemize}

For the deep learning models, no explicit feature engineering is applied. Raw sEMG signals and the corresponding sEMG token sequences are used as two alternative input representations, with dimensions $T \times C$ and $L \times C$, respectively. Both inputs are encoded by a deep model and mapped to action classes through a task-specific classification head. Although the internal architectures differ, all deep learning models are evaluated under the same two input settings to ensure a consistent comparison between raw sEMG signals and sEMG token sequences. ViT-HGR is presented as a representative example to illustrate how this general framework is instantiated in practice. Each input sequence is divided into temporal patches, which are then projected into a latent embedding space by a learnable linear layer. To aggregate global sequence information, a learnable class token \(\mathbf{x}_{\mathrm{cls}} \in \mathbb{R}^D\) is prepended to the sequence, where $D$ is the model dimension. Positional encoding $\mathbf{E}_{\mathrm{pos}} \in \mathbb{R}^{(P+1) \times D}$ is added to retain temporal order information. Here $P$ is the number of patches and $P=256$ and $8$ for sEMG signals and tokens input respectively. Formally, the input to the Transformer encoder is:
$$
\mathbf{Z}_0 = [\mathbf{x}_{\mathrm{cls}}; \mathbf{x}_1; \mathbf{x}_2; \dots; \mathbf{x}_P] + \mathbf{E}_{\mathrm{pos}}
$$
where $\mathbf{x}_i \in \mathbb{R}^D$ is the embedding for the \(i\)-th patch.

The Transformer Encoder is composed of $\mathcal{L}$ blocks. For the $l$-th block, given the input $\mathbf{Z}_{l-1}$, it is processed sequentially through a Multi-Head Attention (MHA) mechanism and a Feed-Forward Network (FFN), both followed by a residual connection and Layer Normalization (LayerNorm):
$$
\mathbf{Y}_l = \mathrm{LayerNorm}\big(\mathbf{Z}_{l-1} + \mathrm{MHA}(\mathbf{Z}_{l-1})\big)
$$
$$
\mathbf{Z}_l = \mathrm{LayerNorm}\big(\mathbf{Y}_l + \mathrm{FFN}(\mathbf{Y}_l)\big)
$$
where $\mathrm{MHA}(\mathbf{Z}) = \mathrm{Concat}(\mathrm{head}_1, \dots, \mathrm{head}_h) \mathbf{W}^O$, and each attention head is computed by $
\mathrm{head}_i = \mathrm{softmax}\left(\frac{\mathbf{Q}_i \mathbf{K}_i^\top}{\sqrt{d_k}}\right) \mathbf{V}_i$. Through this architecture, the Transformer encoder effectively models both temporal dependencies within each muscle channel and inter-muscle relationships, enabling end-to-end action classification. In our implementation, we set $L = 256$ and $T = \left[ (L - 1) \times 25\, \mathrm{ms} + 50\,\mathrm{ms} \right] \times 1269\, \mathrm{Hz} \approx 8192$. The value of $T$ matches the temporal span of the sEMG tokens sequence ($L = 256$) under a 50 ms window with 25 ms overlap at a sampling rate of 1269 Hz, ensuring temporal alignment for fair comparison. Notably, the tokenized representation reduces the input dimension by approximately 96\% compared to the raw sEMG signals (from $T\times C$ to $L\times C$). For the ViT-HGR model, $D$ and $\mathcal{L}$ are set to 64 and 1 respectively. 

The same comparison protocol is applied to the other deep learning models. For each architecture, the sEMG signals and the corresponding token sequences are standardized over the same temporal interval and transformed by the model-specific input module before action classification. Within each model, the data split, training strategy, and evaluation metrics are kept identical across the two input settings, so that performance differences primarily reflect the effect of the input representation. All experiments are conducted using five-fold subject-wise cross-validation to ensure robustness and generalizability.

\subsubsection{Movement Quality Assessment}

To demonstrate the application potential of the proposed sEMG tokens, we conduct a case study on movement quality assessment. In this setting, movement quality is formulated as a sequence similarity problem between a subject’s execution and a correct reference performed by a fitness expert. Specifically, we use the MEQAD dataset, which contains both correct and faulty executions of several actions. Since different subjects may vary in action duration and speed, we adopt Dynamic Time Warping (DTW) to align and compare sequences of differing lengths.

In our setup, each action is represented by a sequence of sEMG tokens reflecting the activation states of $C$ muscles over time. Formally, each action sequence is denoted as:
\begin{equation}
S=\left[s_1^{\left(1\right)},s_1^{\left(2\right)},...,s_1^{\left(C\right)}\right],\left[s_2^{\left(1\right)},s_2^{\left(2\right)},...,s_2^{\left(C\right)}\right],...,\left[s_T^{\left(1\right)},s_T^{\left(2\right)},...,s_T^{\left(C\right)}\right]
\end{equation}
where $s_i^{(m)}$ is the token value of the $m$-th muscle at time step $i$, with $T$ representing the total number of time steps.

To compare two sequences $S^{A}$ and $S^{B}$, we define the pointwise distance between time steps $i$ and 
$j$ as the L1 norm across all muscle channels. The DTW distance is then computed recursively via dynamic programming:
\begin{equation}
D(i,j)=\mathrm{L1}(S_i^{A},S_i^{B})+\min{(D(i-1,j-1),D(i-1,j),D(i,j-1))}
\end{equation}
where $D(i,j)$ denotes the cumulative alignment cost between the first $i$ steps of sequence $A$ and the first $j$ steps of sequence $B$.

The final DTW distance $\mathrm{DTW}(S^A,S^B)=D(L_A,L_B)$, where $L_A,L_B$ are the sequence lengths. It provides a temporal alignment score that reflects the overall similarity in muscle states between two executions. Despite its simplicity, this distance metric captures both timing and pattern differences in a compact form.

To translate the DTW distance into an interpretable quality score, we follow the method in \citep{yu2019dynamic} and compute:
\begin{equation}
\mathrm{Similarity~Score}=1-\frac{\mathrm{DTW}(S^A,S^B)}{\mathrm{Max}_{\mathrm{error}}\times C\times \hat{L}}
\end{equation}
Here, $\mathrm{Max}_{\mathrm{error}}$ denotes the maximum possible token difference (e.g., from rest to maximal contraction), which equals $K-1$ in our setting. $\hat{L}$ is the length of the optimal warping path. The resulting score ranges from 0 to 100\%, with higher values indicating greater similarity to the correct execution.

This approach provides a simple yet effective means of quantifying movement quality based on tokenized sEMG sequences.

\section{Results and Analysis}\label{sec3}
\subsection{Effect of the Number of Muscle-State Units}

The number of muscle-state units $K$ determines the granularity of the resulting discrete representation. A smaller $K$ may partition the sEMG feature space too coarsely to preserve subtle variations in muscle states. In contrast, a larger $K$ enables a finer-grained representation but may introduce subject-specific states, thereby reducing inter-subject consistency. We therefore evaluate the appropriate value of $K$ from two complementary perspectives: consistency and representation capacity.

\begin{figure}
    \centering
    \includegraphics[width=0.75\textwidth]{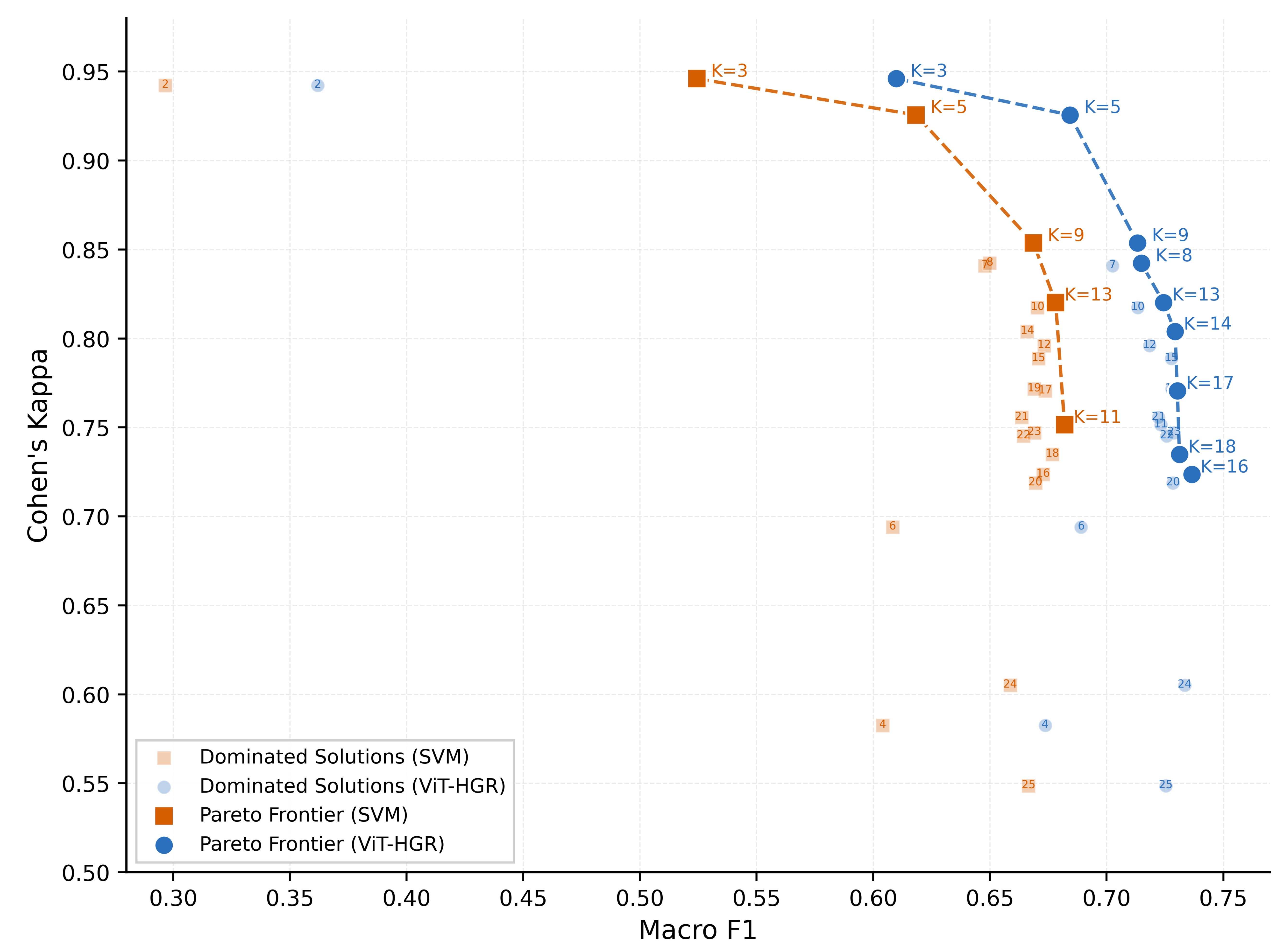}
    \caption{Trade-off between sEMG token consistency and representational capacity across different numbers of muscle-state units. }
    
    \label{Figure_11}
\end{figure}

As shown in Figure~\ref{Figure_11}, smaller numbers of states generally yielded higher token assignment consistency, while the complete results are reported in Table~\ref{tab:k_sensitivity} in the Appendix. For example, Cohen’s Kappa reached 0.94 $\pm$ 0.03, 0.95 $\pm$ 0.03, and 0.93 $\pm$ 0.03 at $K=2$, $K=3$, and $K=5$, respectively. One possible explanation is that coarse-grained partitions cover broader regions of the feature space and may therefore be less sensitive local signal perturbations. However, the consistency did not vary monotonically with $K$ . In particular, Cohen’s Kappa decreased to 0.58 $\pm$ 0.30 and 0.69 $\pm$ 0.26 at $K=4$ and $K=6$, respectively, with substantial variation across folds. These results suggest that partition stability depends not only on the number of states but also on the sensitivity of the resulting partitions to inter-subject variations.

We subsequently evaluated the discriminative information retained by the token representations using a downstream action-recognition task. SVM and ViT-HGR were selected as representative classifiers with different modeling capacities. As $K$  increased from 2, both models showed substantial improvements in classification performance, suggesting that a larger state vocabulary allows the token representation to preserve finer-grained muscle-state information. Specifically, the accuracy of SVM increased from 0.3458 $\pm$ 0.0693 at $K=2$ to a maximum of 0.6954 $\pm$ 0.0380 at $K=11$, whereas ViT-HGR achieved its highest accuracy of 0.7619 $\pm$ 0.0224 at $K=16$. However, the performance gains gradually diminished in the intermediate range, and both models exhibited an approximate performance plateau between $K=11$ and $K=18$. Further increasing $K$  did not yield consistent improvements, suggesting that excessively fine-grained partitions may increase representational complexity without consistently providing additional information relevant to action discrimination.

To jointly evaluate the trade-off between consistency and representational capacity, we constructed separate Pareto frontiers using Cohen’s Kappa and Macro-F1 as two objectives to be maximized. A candidate $K$ was considered non-dominated if no other candidate achieved performance no worse in both objectives and strictly better in at least one objective. As shown in Fig.~\ref{Figure_11}, The shared non-dominated solutions, $K\in \{3, 5, 9, 13\}$, appeared on the Pareto frontiers of both classifiers, indicating that their trade-offs were relatively robust across different modeling paradigms. Since our goal is to construct an sEMG token representation that is both cross-subject stable and sufficiently informative for downstream tasks, $K=13$ was selected because it retained richer action-discriminative information while maintaining high token assignment consistency. It should be noted that $K$=13 is not the unique optimum across all evaluation criteria, but rather a reasonable choice that balances consistency and representational capacity under the current experimental setting.

\subsection{Multi-level Consistency of sEMG Token}
The preceding analysis showed that the token assignments achieved an overall Cohen's Kappa of 0.82. However, this aggregate result does not reveal whether the observed agreement was consistently maintained across every subjects, muscle channels, and action categories. We therefore conducted a multi-level analysis to characterize the distribution and heterogeneity of token assignment consistency.

Table~\ref{tab:multilevel_kappa} summarizes token consistency at the subject, muscle, and action levels. At the subject level, Cohen’s Kappa was 0.8304 $\pm$ 0.0968, with 9 of the 14 subjects achieving Cohen’s Kappa $\geq$ 0.81. This result indicates that the sEMG token was reproducible for most subjects and generalized reasonably well to unseen subject data. At the muscle level, the Cohen’s Kappa was 0.8165 $\pm$ 0.0204, and 12 of the 16 muscles achieved Cohen’s Kappa $\geq$ 0.81. The narrow range of 0.7751-0.8405 indicates that consistency varied only modestly across muscles. This suggests that the tokens was not dominated by a particular muscle channel, but was able to characterize recurring patterns across different muscles. At the action level, the Cohen’s Kappa was 0.8161 $\pm$ 0.0169, with 26 of the 43 actions achieving Cohen’s Kappa $\geq$ 0.81. The limited variation across actions indicates that the tokens were not specific to individual action categories, but instead represented local muscle activities that could be reused across different movements. 

Taken together, these results suggest that the tokens represent stable patterns that can be reproducibly identified across subjects, muscles, and actions.

\begin{table}[t]
\centering
\caption{Cohen's Kappa statistics for token consistency at different analysis levels.}
\label{tab:multilevel_kappa}
\renewcommand{\arraystretch}{1.15}
\setlength{\tabcolsep}{6pt}
\begin{tabular}{lccccc}
\toprule
\textbf{Analysis level}
& \textbf{$N$}
& \textbf{Mean $\pm$ SD}
& \textbf{Median}
& \textbf{Range}
& \textbf{$ \text{Cohen's Kappa} \geq 0.81$} \\
\midrule
Subject & 14 & $0.8304 \pm 0.0968$ & $0.8675$ & $[0.6729$--$0.9216]$ & 9/14 \\
Muscle  & 16 & $0.8165 \pm 0.0204$ & $0.8247$ & $[0.7751$--$0.8405]$ & 12/16 \\
Action  & 43 & $0.8161 \pm 0.0169$ & $0.8170$ & $[0.7805$--$0.8463]$ & 26/43 \\
\bottomrule
\end{tabular}
\end{table}

\subsection{Interpretability of sEMG Tokens}\label{Interpretability}

\begin{figure}
    \centering
    \includegraphics[width=0.99\textwidth]{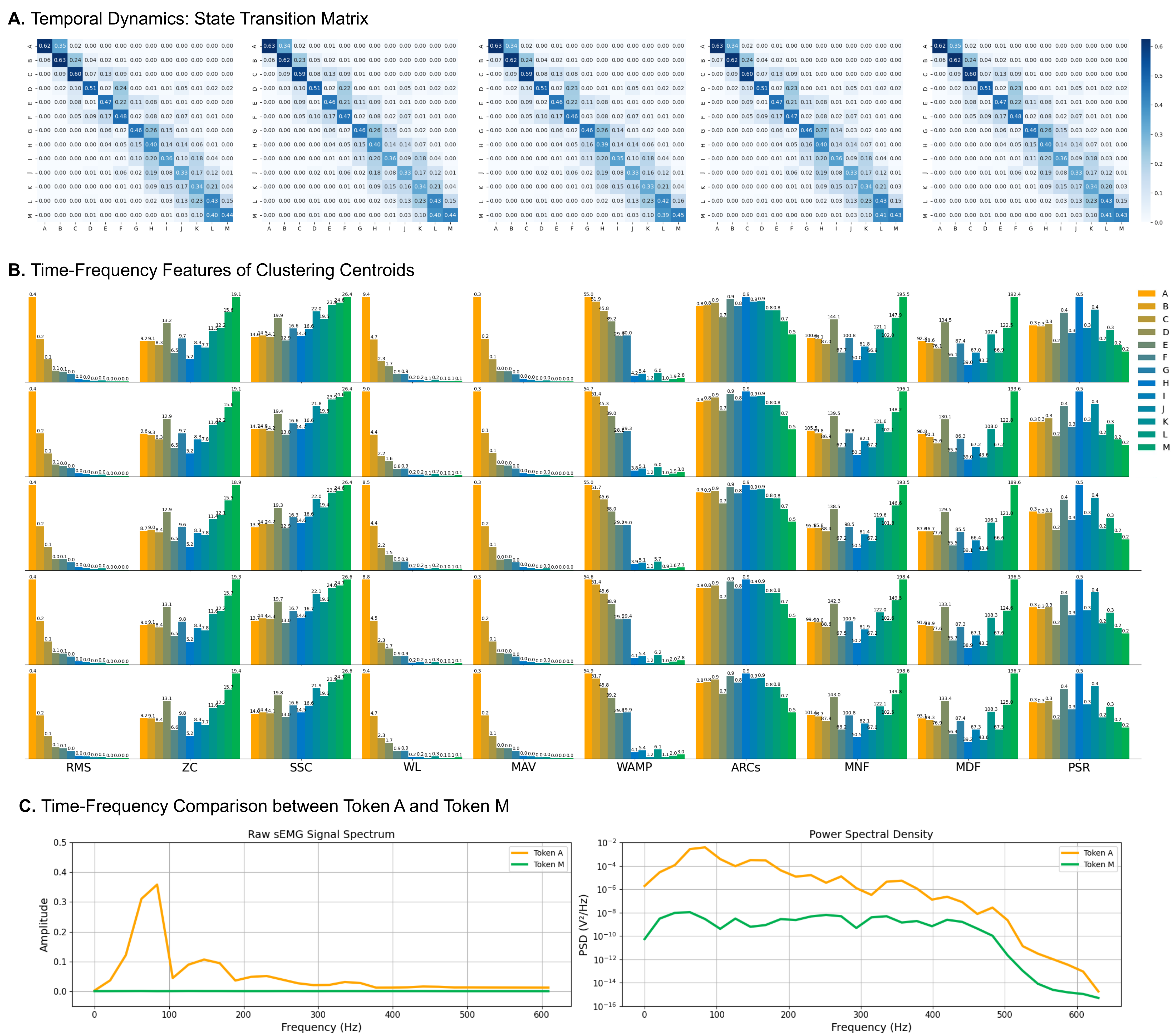}
    \caption{Interpretability of sEMG tokens. 
    (A) Token transition matrices reveal the temporal evolution of the token sequences.
    (B) Distribution of cluster centroid features across folds reflects consistent representations of muscle states.
    (C) Power spectral comparison between token A and token M.}
    \label{Physiological Structuring of sEMG Tokens}
\end{figure}

To investigate whether the sEMG tokens exhibit an intrinsic organization rather than forming an arbitrary set of discrete states, we first examined their temporal transition patterns. Token-level transition probability matrices are constructed to characterize the temporal organization of the token sequences. As shown in Fig~\ref{Physiological Structuring of sEMG Tokens}-A, the results across five-fold cross-validation exhibit similar transition patterns, with transitions concentrated same identical or closely related tokens. This pattern suggests that token in adjacent time windows tend to evolve coherently rather than switching randomly, which is consistent with the progressive temporal evolution of muscle activity. We further present the feature vectors of the cluster centroids in Fig~\ref{Physiological Structuring of sEMG Tokens}-B to investigate the potential physiological interpretability of the generated sEMG tokens. Notably, across all five-fold cross-validation experiments, token $A$ consistently exhibits relatively high values for several representative time-domain features, including RMS, WL, MAV, and WAMP, which are commonly associated with the intensity of muscle activity. According to the motor unit recruitment theory (Henneman’s size principle) \citep{henneman1965functional,farina2014extraction}, stronger contractions generally involve the recruitment of additional and larger motor units, resulting in increased sEMG amplitude. Therefore, we infer that token $A$ is associated with a state of strong muscle contraction, likely reflecting periods of active effort or peak activation. In contrast, token $M$ exhibits the lowest values for these amplitude-related features and relatively small signal fluctuations. Further spectral analysis reveals that, as shown in Fig.~\ref{Physiological Structuring of sEMG Tokens}-C, token $A$ exhibits concentrated spectral energy around 100 Hz, whereas token $M$ shows a flatter and lower-energy spectral profile that is closer to the recording baseline \citep{boyer2023reducing,DELUCA20101573}. These observations suggest that token $M$ may be associated with a resting or minimally activated muscle activity, whereas token $A$ may be associated with relatively strong muscle activity. The remaining tokens may represent a range of muscle states between these two extremes, reflecting gradual variations in muscle activation intensity.

To further evaluate the interpretability of the sEMG tokens, we examined whether their occupancy distributions were related to the functional involvement of muscles during actual movements. During human movement, agonist muscles act as the primary force-producing muscles responsible for generating the joint torques required to execute an action and therefore generally exhibit stronger activation than muscles that are not primarily involved in force production. Based on this biomechanical prior, we hypothesize that, if the sEMG tokens effectively map distinct muscle activity states, their occupancy distributions should vary systematically according to the functional roles of the muscles, such as whether a muscle serves as an agonist or a non-agonist. To examine this hypothesis, we selected five representative actions involving relatively simple movement patterns and a limited number of agonist muscles. For each action, we compared the token occupancy distribution of the corresponding agonist muscle with the aggregated distribution of the non-agonist muscles. As shown in Table~\ref{tab:token_distribution_clean}, clear and consistent differences can be observed between the two functional muscle groups. Across the five actions, tokens A--G account for approximately 74.9\% of the assignments in the agonist muscles but only 15.6\% in the non-agonist muscles. Conversely, tokens H--M account for 25.1\% and 84.4\% of the assignments in the agonist and non-agonist muscles, respectively. For example, during Seated Toe Raises, tokens A--D account for 66.9\% of the assignments in the tibialis anterior, whereas they are almost absent from the non-agonist muscles. In contrast, tokens K--M account for 63.1\% of the assignments in the non-agonist muscles. These consistent distributional shifts indicate that tokens are associated with the functional involvement of muscles during movement rather than being arbitrarily distributed. 

Together with the centroid and spectral analyses, these results support the interpretability of the tokens. Although these analysis does not provide a direct physiological validation of each token, it provides preliminary evidence that the token representation has interpretable structure and can reflect meaningful differences in muscle activation patterns.



\begin{table*}[t]

\centering

\caption{Token occupancy distributions of agonist and non-agonist muscle groups across representative actions. Values are reported as percentages.}

\label{tab:token_distribution_clean}

\scriptsize

\setlength{\tabcolsep}{3.8pt}

\renewcommand{\arraystretch}{1.12}

\begin{tabular}{l l l *{13}{r}}

\toprule

\textbf{Action Name}
& \textbf{Muscle}
& \multicolumn{13}{c}{\textbf{Token occupancy ratio (\%)}} \\

\cmidrule(lr){3-15}

& 
& \textbf{A} & \textbf{B} & \textbf{C} & \textbf{D} & \textbf{E} & \textbf{F} & \textbf{G}
& \textbf{H} & \textbf{I} & \textbf{J} & \textbf{K} & \textbf{L} & \textbf{M} \\

\midrule

Seated Toe Raises
& Tibialis anterior (Agonist)
& 9.6 & 23.1 & 14.8 & 19.4 & 1.5 & 6.2 & 1.3 & 2.6 & 3.1 & 3.9 & 5.2 & 6.2 & 3.1 \\

& Non-agonist
& 0.0 & 0.0 & 0.0 & 0.1 & 0.9 & 1.4 & 3.4 & 11.2 & 9.3 & 10.6 & 20.0 & 28.0 & 15.1 \\

\midrule

Biceps Curl
& Biceps brachii (Agonist)
& 1.4 & 3.2 & 10.2 & 1.0 & 21.2 & 10.6 & 12.9 & 17.4 & 5.9 & 5.5 & 5.1 & 4.4 & 1.3 \\

& Non-agonist
& 0.0 & 0.1 & 0.8 & 1.0 & 1.5 & 3.6 & 3.0 & 10.2 & 8.0 & 13.2 & 20.0 & 25.9 & 12.6 \\

\midrule

Arm Circles
& Deltoid (Agonist)
& 4.9 & 15.6 & 25.6 & 4.7 & 17.8 & 10.0 & 9.4 & 6.7 & 2.8 & 1.3 & 0.9 & 0.3 & 0.1 \\

& Non-agonist
& 0.7 & 2.0 & 4.5 & 2.3 & 5.1 & 6.2 & 3.5 & 9.2 & 6.2 & 12.3 & 14.9 & 22.3 & 11.0 \\

\midrule

Standing Overhead Press
& Deltoid (Agonist)
& 5.9 & 20.4 & 28.2 & 5.4 & 14.0 & 12.5 & 4.8 & 5.4 & 1.2 & 1.5 & 0.4 & 0.2 & 0.0 \\

& Non-agonist
& 0.2 & 1.8 & 5.1 & 3.2 & 4.9 & 7.9 & 4.7 & 8.8 & 7.4 & 10.8 & 16.0 & 20.2 & 9.1 \\

\midrule

Calf Raises
& Gastrocnemius lateral head (Agonist)
& 0.0 & 2.2 & 9.9 & 34.0 & 1.0 & 11.4 & 0.5 & 3.5 & 1.8 & 13.0 & 5.0 & 11.9 & 5.9 \\

& Non-agonist
& 0.0 & 0.0 & 0.2 & 1.7 & 1.2 & 4.3 & 2.9 & 10.7 & 9.1 & 17.0 & 21.3 & 23.9 & 7.7 \\

\bottomrule

\end{tabular}

\vspace{2pt}

\begin{minipage}{0.98\textwidth}

\footnotesize

\textit{Note:} Agonist muscles denote primary force-producing muscles, while non-agonist muscles refer to all remaining recorded muscles. Tokens A--M correspond to 13 token.

\end{minipage}

\end{table*}

\subsection{Comparison of Human Action Recognition}

To investigate whether the sEMG token representation mitigates the impact of inter-subject variability, we compared its performance with that of sEMG signals in the cross-subject human action recognition setting. Unlike the preceding analysis, which focused on the effect of token granularity, this experiment directly evaluates the downstream utility of the selected sEMG token representation relative to the sEMG signals. 

\subsubsection{Comparison with Raw sEMG Signals}

We first compared sEMG signals and discrete sEMG tokens across five representative classifiers using the same cross-subject data splits and corresponding model-specific training protocols. As shown in Table~\ref{tab:token_signal_comparison}, the sEMG token representation improved accuracy, Macro-P, Macro-R, and Macro-F1 across all five models. The improvements were particularly pronounced for the conventional machine-learning classifiers. Compared with raw sEMG signals, the token representation improved the accuracy of SVM and KNN by 3.50\% and 11.38\%, respectively, while their Macro-F1 scores increased by 6.34\% and 13.58\%. One possible explanation for the larger gains observed with KNN is that discretization may reduce the influence of local signal variability and provides a more structured representation for distance-based comparison. This may alleviate some of the difficulties encountered when directly comparing raw sEMG signals affected by inter-subject variability. The three deep-learning models also consistently benefited from the token representation. The accuracy increased by 2.73\%, 5.21\%, and 3.42\%  for ViT-HGR, WaveFormer, and STCNet, respectively, while the corresponding Macro-F1 gains were 3.18\%, 4.29\%, and 4.42\%. These results show that the benefits of sEMG tokens are not limited to conventional classifiers with limited representation-learning capacity. Even when used with deep models capable of learning complex temporal and task-specific representations, the discrete sEMG tokens yielded better accuracy and macro-averaged performance than the raw signals. 

\begin{figure}
    \centering
    \includegraphics[width=1\linewidth]{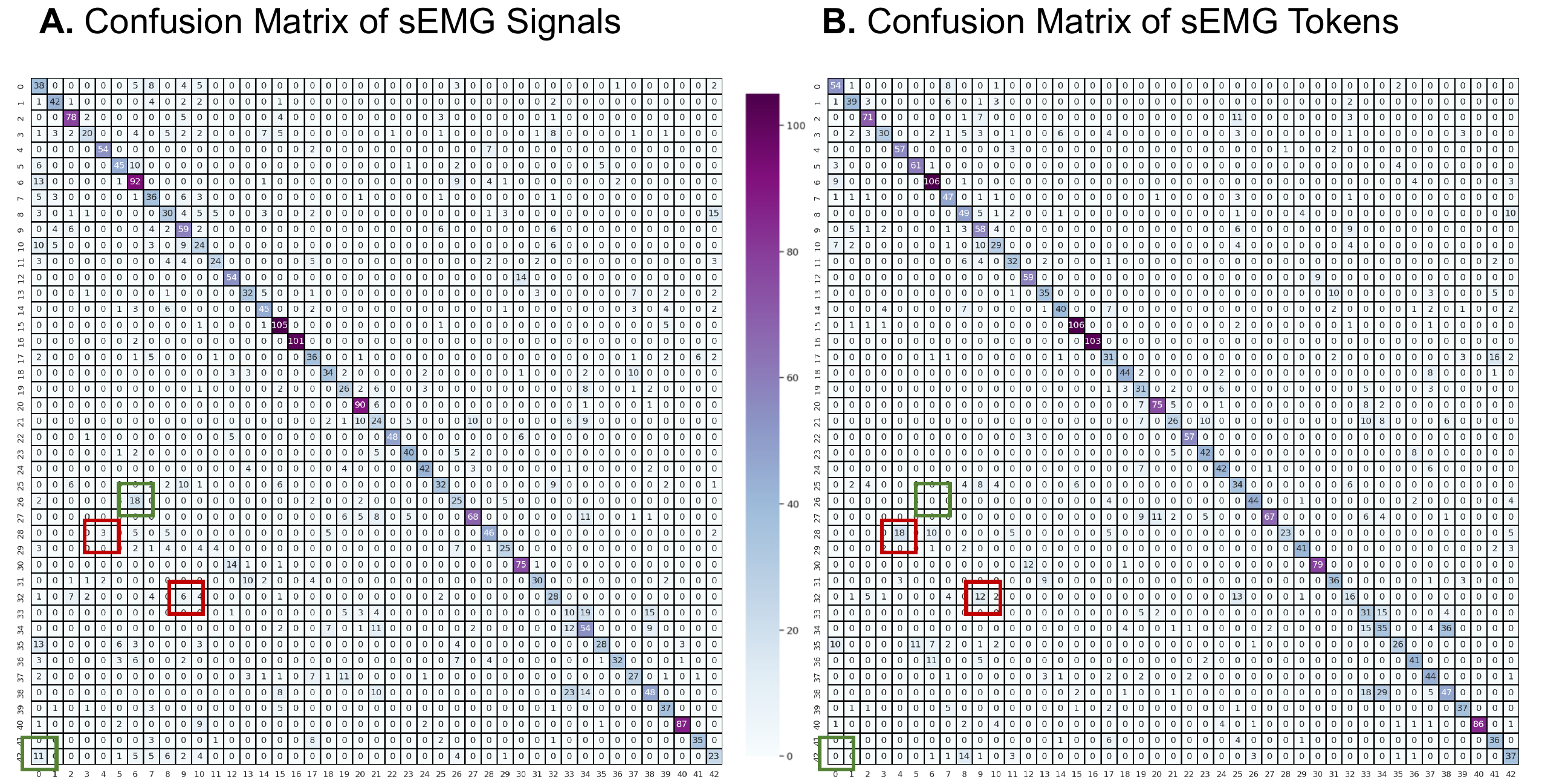}
    \caption{The aggregated confusion matrices of the SVM model across five-fold cross-validation. (A) shows the confusion matrix based on raw sEMG signals; (B) shows the confusion matrix based on sEMG tokens. In each confusion matrix, the vertical axis represents the ground truth labels and the horizontal axis represents the predicted labels. The action labels correspond to all indices are provided in Appendix Table.~\ref{ActionEMG-43}. Green boxes highlight several misclassifications corrected by the sEMG tokens, while red boxes indicate some new misclassifications introduced by the token-based method. }
    \label{signalvstoken}
\end{figure}

Taking the SVM model as an example, we further examined the confusion matrices to investigate how the use of sEMG tokens affected recognition errors. As shown in Fig.~\ref{signalvstoken}, several misclassifications were reduced when the models were trained using sEMG tokens. For example, the number of misclassifications between Single-Leg Stand (class 26) and Sit Up (class 6) decreased to 0 cases, while that between Standing External Hip Rotation (class 42) and Jogging (class 0) decreased to 0 cases. These reductions suggest that the token representation captures the dominant muscle states of these actions more effectively, thereby improving their separability. However, the sEMG tokens also led to several new misclassifications, primarily between actions with highly similar muscle states. For instance, misclassifications between Horse Stance (class 28) and Squats (class 4), as well as between Bounding (class 32) and Lunge Jumps (class 25), increased modestly. These misclassifications may arise because the paired actions recruit similar muscles and exhibit comparable activation timing and intensity. From another perspective, this behavior also suggests that the token representation captures the muscle-state dynamics shared by these actions. While such an abstraction improves robustness by emphasizing task-relevant muscle states, it may also reduce sensitivity to subtle within-state variations that are needed to distinguish actions with highly similar muscle states.

The direct comparison between raw sEMG signals and sEMG tokens demonstrates that the token representation provides superior performance in the cross-subject action recognition setting. These results suggest that, unlike raw continuous sEMG streams which are highly susceptible to individual physiological idiosyncrasies, sEMG tokens discretize ‌high-frequency signals into compact, structured sequences of muscle states. This formulation effectively preserves action-essential muscle activation profiles while mitigates the adverse impacts of subject-specific variations on downstream performance. Consequently, the proposed sEMG tokens serve as an efficient, more robust alternative representation to continuous raw sEMG signals.

\subsubsection{Comparison with Pretrained sEMG Representations}

To further examine whether the performance gains of sEMG tokens were merely due to the additional feature-extraction step relative to raw sEMG signals, we conduct additional experiments based on the STCNet framework. In the first stage of STCNet, the encoder is pretrained on raw sEMG signals using a subject-aware contrastive learning (SAC) module, encouraging it to learn latent representations that are less sensitive to inter-subject variability. Building on this pretrained backbone, we compare two input representations under an identical classifier architecture to ensure a fair evaluation: (1) the latent representations produced by the first stage of STCNet, and (2) the sEMG token representations proposed in this work. 

As shown in Table~\ref{tab:token_latent_comparison}, the proposed sEMG tokens achieved better performance than the SAC-pretrained representations in most metrics. Specifically, sEMG tokens improved accuracy from 0.7570 to 0.7802, corresponding to an absolute gain of 2.31\%. The macro-averaged metrics also improved consistently, with Macro-P, Macro-R, and Macro-F1 increasing by 3.53\%, 1.79\%, and 1.80\% , respectively. 

The above results show that even when raw sEMG signals are encoded into high-level representations by a pretrained model, sEMG tokens still achieve better performance. This finding indicates that the advantage of sEMG tokens cannot be attributed solely to the additional front-end feature extraction step. Rather, it suggests that the discrete token representation itself contributes to a more robust cross-subject representation of muscle activity.





\begin{table*}[t]
\centering
\caption{Action-recognition performance of raw sEMG signals and discrete
sEMG tokens across conventional machine-learning and deep-learning models.
Results are reported as mean $\pm$ standard deviation over five cross-subject
folds.}
\label{tab:token_signal_comparison}

\setlength{\tabcolsep}{5.0pt}
\renewcommand{\arraystretch}{1.08}
\scriptsize

\resizebox{\textwidth}{!}{%
\begin{tabular}{l c cccc}
\toprule
\multirow{2}{*}{Model}
& \multirow{2}{*}{Input}
& \multicolumn{4}{c}{Recognition Performance} \\
\cmidrule(lr){3-6}
&
& Accuracy
& Macro-P
& Macro-R
& Macro-F1 \\
\midrule

\multicolumn{6}{l}{\textit{Conventional machine-learning models}} \\
\addlinespace[1pt]

\multirow{3}{*}{SVM}
& signals
& $0.6440 \pm 0.0430$
& $0.6747 \pm 0.0388$
& $0.6294 \pm 0.0374$
& $0.6148 \pm 0.0404$ \\

& tokens
& $0.6790 \pm 0.0339$
& $0.7188 \pm 0.0394$
& $0.6933 \pm 0.0311$
& $0.6782 \pm 0.0323$ \\

& $\Delta$ (pp)
& $\mathbf{+3.50}$ **
& $\mathbf{+4.41}$ *
& $\mathbf{+6.39}$ **
& $\mathbf{+6.34}$ ** \\
\hdashline
\addlinespace[3pt]

\multirow{3}{*}{KNN}
& signals
& $0.4333 \pm 0.0153$
& $0.5270 \pm 0.0445$
& $0.4085 \pm 0.0182$
& $0.3909 \pm 0.0304$ \\

& tokens
& $0.5471 \pm 0.0524$
& $0.6139 \pm 0.0464$
& $0.5531 \pm 0.0414$
& $0.5267 \pm 0.0470$ \\

& $\Delta$ (pp)
& $\mathbf{+11.38}$ ***
& $\mathbf{+8.69}$  ***
& $\mathbf{+14.46}$ ***
& $\mathbf{+13.58}$  ***\\

\midrule
\addlinespace[2pt]
\multicolumn{6}{l}{\textit{Deep learning models}} \\
\addlinespace[1pt]

\multirow{3}{*}{ViT-HGR}
& signals
& $0.7277 \pm 0.0240$
& $0.7405 \pm 0.0276$
& $0.7085 \pm 0.0205$
& $0.6926 \pm 0.0216$ \\

& tokens
& $0.7550 \pm 0.0248$
& $0.7620 \pm 0.0309$
& $0.7386 \pm 0.0227$
& $0.7244 \pm 0.0248$ \\

& $\Delta$ (pp)
& $\mathbf{+2.73}$ *
& $\mathbf{+2.15}$ 
& $\mathbf{+3.01}$ *
& $\mathbf{+3.18}$ * \\
\hdashline
\addlinespace[3pt]

\multirow{3}{*}{WaveFormer}
& signals
& $0.7366 \pm 0.0530$
& $0.7424 \pm 0.0419$
& $0.7280 \pm 0.0484$
& $0.7182 \pm 0.0468$ \\

& tokens
& $0.7887 \pm 0.0368$
& $0.7964 \pm 0.0396$
& $0.7737 \pm 0.0416$
& $0.7611 \pm 0.0426$ \\

& $\Delta$ (pp)
& $\mathbf{+5.21}$ **
& $\mathbf{+5.40}$ **
& $\mathbf{+4.57}$ **
& $\mathbf{+4.29}$ **\\
\hdashline
\addlinespace[3pt]

\multirow{3}{*}{STCNet}
& signals
& $0.7570 \pm 0.0462$
& $0.7688 \pm 0.0592$
& $0.7459 \pm 0.0487$
& $0.7390 \pm 0.0539$ \\

& tokens
& $0.7912 \pm 0.0349$
& $0.8236 \pm 0.0261$
& $0.7912 \pm 0.0352$
& $0.7832 \pm 0.0327$ \\

& $\Delta$ (pp)
& $\mathbf{+3.42}$ *
& $\mathbf{+5.48}$ *
& $\mathbf{+4.53}$ **
& $\mathbf{+4.42}$ **\\

\bottomrule
\end{tabular}%
}

\vspace{3pt}
\begin{minipage}{0.98\textwidth}
\footnotesize
\textit{Note:}
$\Delta$ denotes the absolute performance difference between the token and
signal representations, reported in percentage points (pp) and calculated as
$\Delta=(\mathrm{Token}-\mathrm{Signal})\times100$.
Positive performance gains are shown in bold.  With $p$ as the p-value of the paired t-test, *$p$ < 0.1, **$p$ < 0.05, ***$p$ < 0.01 
\end{minipage}

\end{table*}

\begin{table*}[t]
\centering
\caption{Comparison of SAC-pretrained sEMG representations and sEMG tokens using the same downstream classifier.
Results are reported as mean $\pm$ standard deviation over five cross-subject
folds.}
\label{tab:token_latent_comparison}

\setlength{\tabcolsep}{5.0pt}
\renewcommand{\arraystretch}{1.08}
\scriptsize

\resizebox{\textwidth}{!}{%
\begin{tabular}{m{2.5cm} c cccc}
\toprule
\multirow{2}{*}{Model}
& \multirow{2}{*}{Input}
& \multicolumn{4}{c}{Recognition Performance} \\
\cmidrule(lr){3-6}
&
& Accuracy
& Macro-P
& Macro-R
& Macro-F1 \\
\midrule

\addlinespace[3pt]

\multirow{3}{*}{STCNet(w/o SAC)}

& {\tiny \makecell[l]{SAC-pretrained\\representations}}
& $0.7570 \pm 0.0462$
& $0.7688 \pm 0.0592$
& $0.7459 \pm 0.0487$
& $0.7390 \pm 0.0539$ \\
\addlinespace[3pt]

& tokens
& $0.7802 \pm 0.0410 $
& $0.8041 \pm 0.0301$
& $0.7638 \pm 0.0393$
& $0.7570 \pm 0.0382$ \\

\addlinespace[3pt]

& $\Delta$ (pp)
& $\mathbf{+2.31}$
& $\mathbf{+3.53}$
& $\mathbf{+1.79}$
& $\mathbf{+1.80}$ \\

\bottomrule
\end{tabular}%
}

\vspace{3pt}
\begin{minipage}{0.98\textwidth}
\footnotesize
\textit{Note:}
STCNet denotes the complete two-stage framework, in which the model is first pretrained using subject-aware contrastive learning (SAC) to learn subject-invariant representations and is then fine-tuned for action classification. STCNet (w/o SAC) denotes the variant without contrastive pretraining and serves as a classifier.

\end{minipage}
\end{table*}

\subsection{Ablation Study}

To examine the contribution of each feature group to sEMG tokenization, we performed a group-wise ablation study in which one feature category was removed at a time. Cohen’s kappa was used to quantify the consistency of token assignments, while Macro-F1 scores obtained from five classifiers were used to evaluate the discriminative information preserved in the resulting token representations. The selected features were grouped according to the signal properties they capture: amplitude and waveform magnitude (RMS, MAV, and WL), waveform complexity and variation (ZC, SSC, and WAMP), spectral distribution (MNF, MDF, and PSR), and autoregressive dynamics (ARCs). Together, these feature groups characterize complementary aspects of sEMG, including signal magnitude, local waveform fluctuations, spectral energy distribution, and short-term temporal dependencies.

As shown in Table~\ref{tab:feature_ablation}, the feature-free baseline, which directly used windowed raw sEMG signals without extracting the selected handcrafted features, yielded substantially weaker results. Its Cohen’s kappa was only $0.40$, indicating poor consistency of the resulting tokens across folds. In addition, this baseline achieved the loweer Macro-F1 scores across all five classifiers, with particularly large gaps compared with the complete feature set for SVM, KNN, and STCNet. These results suggest that directly discretizing raw sEMG windows is insufficient for obtaining stable and discriminative muscle-state tokens. Instead, the selected feature groups provide an essential feature space for constructing token representations. After confirming the necessity of feature-based tokenization, we further examined the contribution of each feature group by removing one group at a time. removing "RMS\&MAV\&WL" slightly increased Cohen’s kappa from $0.82$ to $0.84$, but reduced the Macro-F1 scores of all five classifiers, with decreases ranging from $0.0103$ to $0.0479$. These results indicate that amplitude- and magnitude-related features preserve information relevant to action discrimination, although their contribution to token assignment consistency appears limited. Removing ZC, SSC, and WAMP produced the largest and most consistent degradation in downstream performance. The Macro-F1 score decreased for every classifier, including reductions of $0.1110$ for WaveFormer and $0.1571$ for STCNet, whereas Cohen’s kappa decreased only slightly from $0.82$ to $0.81$. These findings indicate that waveform-complexity features make a substantial contribution to the discriminative capability of the token representation by capturing local oscillations, slope reversals, and significant sample-to-sample variations in the sEMG signal. In contrast, removing "MNF\&MDF\&PSR" reduced Cohen’s kappa from $0.82$ to $0.77$, while its effect on downstream classification varied across model architectures. SVM and KNN showed slightly lower Macro-F1 scores, whereas ViT-HGR, WaveFormer, and STCNet achieved higher scores. Thus, spectral features appear to contribute more clearly to token assignment consistency, while their contribution to downstream discrimination is model dependent. Removing ARCs similarly reduced Cohen’s kappa, from $0.82$ to $0.78$. SVM and KNN showed modest improvements, whereas the three deep-learning classifiers exhibited lower Macro-F1 scores. This pattern suggests that ARCs contribute to token consistency and may provide complementary local temporal information for deep-learning-based classification.

Overall, the four feature groups contributed complementary information to sEMG tokenization. Amplitude-related and waveform-complexity features mainly supported downstream discrimination, whereas spectral features and ARCs contributed more strongly to token consistency. The complete feature set achieved the lowest average rank across the evaluated metrics and was therefore retained as the final configuration.

\begin{table*}[t]
\centering
\caption{Group-wise ablation results for the selected sEMG features. "w/o" indicates the removal of the corresponding feature group. The best and second-best results are shown in bold and underlined, respectively.}
\label{tab:feature_ablation}
\renewcommand{\arraystretch}{1.15}
\setlength{\tabcolsep}{5.0pt}
\small
\resizebox{\columnwidth}{!}{
\begin{tabular}{lccccccc}
\toprule
\multirow{2}{*}{\textbf{Feature configuration}}
& \multirow{2}{*}{\textbf{Cohen’s kappa}}
& \multicolumn{5}{c}{\textbf{Macro-F1}}
& \multirow{2}{*}{\textbf{Avg. Rank}} \\
\cmidrule(lr){3-7}
& & \textbf{SVM} & \textbf{KNN} & \textbf{ViT-HGR}
& \textbf{WaveFormer} & \textbf{STCNet} & \\

\midrule
only sEMG signal
& $0.40 \pm 0.19$
& $0.5121 \pm 0.0185$
& $0.2352 \pm 0.0191$
& $0.6488 \pm 0.0396$
& $0.6934 \pm 0.0382$
& $0.6574 \pm 0.0411$
& $5.50$ \\
\midrule

w/o RMS\&MAV\&WL
& $\mathbf{0.84 \pm 0.07}$
& $0.6307 \pm 0.0508$
& $0.4788 \pm 0.0492$
& $0.7141 \pm 0.0106$
& $0.7460 \pm 0.0327$
& $0.7399 \pm 0.0316$
& $3.50$ \\

w/o ZC\&SSC\&WAMP
& $0.81 \pm 0.07$
& $0.6447 \pm 0.0480$
& $0.4341 \pm 0.0391$
& $0.6447 \pm 0.0247$
& $0.6501 \pm 0.0497$
& $0.6261 \pm 0.0588$
& $5.00$ \\

w/o MNF\&MDF\&PSR
& $0.77 \pm 0.05$
& $0.6746 \pm 0.0393$
& $0.5159 \pm 0.0436$
& $\mathbf{0.7681 \pm 0.0282}$
& $\mathbf{0.7790 \pm 0.0285}$
& $\mathbf{0.7877 \pm 0.0261}$
& $\underline{2.33}$ \\

w/o ARCs
& $0.78 \pm 0.10$
& $\mathbf{0.6906 \pm 0.0394}$
& $\mathbf{0.5403 \pm 0.0433}$
& $0.7208 \pm 0.0268$
& $0.7459 \pm 0.0300$
& $0.7560 \pm 0.0374$
& $2.67$ \\

\midrule
\rowcolor{gray!12}
\textbf{All features}
& $\underline{0.82 \pm 0.09}$
& $\underline{0.6782 \pm 0.0323}$
& $\underline{0.5267 \pm 0.0470}$
& $\underline{0.7244 \pm 0.0248}$
& $\underline{0.7611 \pm 0.0426}$
& $\underline{0.7832 \pm 0.0327}$
& $\mathbf{2.00}$ \\

\bottomrule
\end{tabular}
}
\end{table*}

\subsection{Sensitivity Analysis of Window Length}\label{window-length}

To examine the effect of analysis window length on sEMG tokenization, we compared five window lengths, including 25, 50, 75, 100, and 200 ms. As shown in Table~\ref{tab:window_length_analysis}, the 50 ms window achieved the best overall performance, with the lowest average rank of 1.17 across Cohen’s kappa and the Macro-F1 scores of the five classifiers. 

In terms of token-assignment consistency, both the 25 ms and 50 ms windows achieved the highest Cohen’s kappa value of 0.82, whereas longer windows showed lower consistency. Specifically, the kappa values decreased to 0.75, 0.74, and 0.64 for the 75, 100, and 200 ms windows, respectively. This trend indicates that relatively short analysis windows are more favorable for generating consistent token assignments across cross-subject folds. In contrast, larger windows may span multiple short-term muscle activation patterns within a single segment, leading to mixed muscle-state information and reducing the specificity of token assignments. For downstream action recognition, the classification performance generally improved when the window length increased from 25 ms to moderate lengths, but declined when the window length was further increased to 200 ms. This suggests that very short windows may provide insufficient samples for stable feature estimation, whereas excessively long windows may smooth transient muscle-activation changes and dilute local muscle-state information. 

Overall, the 50 ms window achieved the best average rank, indicating the most favorable balance between token consistency and downstream discriminative performance.

\begin{table*}[t]
\centering
\caption{Effect of window length on sEMG tokenization. Results are reported as mean $\pm$ standard deviation over five cross-subject folds. The best and second-best results are shown in bold and underlined, respectively. For Avg. Rank, a lower value indicates better overall performance.}
\label{tab:window_length_analysis}

\renewcommand{\arraystretch}{1.15}
\setlength{\tabcolsep}{5.0pt}
\small

\resizebox{\textwidth}{!}{
\begin{tabular}{lccccccc}
\toprule
\multirow{2}{*}{\textbf{Window length}}
& \multirow{2}{*}{\textbf{Cohen's kappa}}
& \multicolumn{5}{c}{\textbf{Macro-F1}} 
& \multirow{2}{*}{\textbf{Avg. Rank}} \\
\cmidrule(lr){3-7}
& 
& \textbf{SVM} 
& \textbf{KNN} 
& \textbf{ViT-HGR}
& \textbf{WaveFormer} 
& \textbf{STCNet} 
& \\

\midrule

25 ms
& $\mathbf{0.82 \pm 0.06}$
& $0.6500 \pm 0.0254$
& $0.4905 \pm 0.0503$
& $0.6932 \pm 0.0346$
& $0.7455 \pm 0.0208$
& $0.7552 \pm 0.0502$
& $4.17$ \\

75 ms
& $\underline{0.75 \pm 0.12}$
& $0.6738 \pm 0.0279$
& $\underline{0.5245 \pm 0.0418}$
& $\mathbf{0.7273 \pm 0.0275}$
& $\underline{0.7594 \pm 0.0174}$
& $0.7634 \pm 0.0294$
& $\underline{2.17}$ \\

100 ms
& $0.74 \pm 0.12$
& $\underline{0.6754 \pm 0.0159}$
& $0.5203 \pm 0.0519$
& $0.7218 \pm 0.0292$
& $0.7521 \pm 0.0242$
& $\underline{0.7661 \pm 0.0359}$
& $2.67$ \\

200 ms
& $0.64 \pm 0.13$
& $0.6576 \pm 0.0276$
& $0.4919 \pm 0.0423$
& $0.7113 \pm 0.0249$
& $0.7459 \pm 0.0376$
& $0.7550 \pm 0.0170$
& $4.17$ \\

\midrule
\rowcolor{gray!12}
\textbf{50 ms}
& $\mathbf{0.82 \pm 0.09}$
& $\mathbf{0.6782 \pm 0.0323}$
& $\mathbf{0.5267 \pm 0.0470}$
& $\underline{0.7244 \pm 0.0248}$
& $\mathbf{0.7611 \pm 0.0426}$
& $\mathbf{0.7832 \pm 0.0327}$
& $\mathbf{1.17}$ \\

\bottomrule
\end{tabular}
}
\end{table*}

\subsection{Inference Efficiency}
The continuous sEMG signal had an input size of $8192 \times 16$, whereas the sEMG token representation reduced the input size to $256 \times 16$. This corresponds to a 32-fold reduction in temporal sequence length and a 96.88\% reduction in the number of input elements.To determine whether the compact token representation translates into improved computational efficiency, we compared the per-sample inference time of continuous sEMG signals and discrete sEMG tokens across five classifiers. As shown in Table~\ref{tab:token_signal_inference}, token-based inputs consistently reduced inference time, although the magnitude of the reduction varied considerably across model types.

The improvement was particularly pronounced for the conventional machine-learning models. For SVM, the inference time decreased from $49.67$ to $4.43$ ms per sample, corresponding to a reduction of approximately $91.08\%$ and an $11.21\times$ speedup. For KNN, the inference time decreased from $15.23$ to $0.74$ ms per sample, representing a reduction of approximately $95.14\%$ and a $20.58\times$ speedup. These substantial gains indicate that the lower-dimensional token representation markedly reduces the computational cost associated with decision-function evaluation and distance-based neighbor search. By contrast, the reductions observed for the deep-learning models were more modest. The inference time decreased by approximately $4.21\%$ for ViT-HGR, $12.78\%$ for WaveFormer, and $1.83\%$ for STCNet. This difference can be attributed to the fact that deep-model inference is determined not only by the raw input length but also by architectural factors such as patch construction, embedding dimensionality, convolutional or attention operations. In our implementation, the signal- and token-based models used comparable downstream architectures and settings to ensure a fair comparison of representation quality. Consequently, much of the backbone computation remained unchanged, limiting the latency reduction achieved solely through input compression.

Overall, sEMG tokens yielded substantial inference acceleration for conventional machine-learning classifiers and modest but consistent reductions for deep-learning models.


\begin{table}[t]
\centering
\caption{Inference efficiency of continuous sEMG signals and discrete sEMG
tokens across different classifiers.}
\label{tab:token_signal_inference}

\setlength{\tabcolsep}{8.0pt}
\renewcommand{\arraystretch}{1.10}
\small

\begin{tabular}{l cc}
\toprule
\multirow{2}{*}{Model}
& \multicolumn{2}{c}{Inference Time (ms/sample)} \\
\cmidrule(lr){2-3}
& Signals & Tokens \\
\midrule

\multicolumn{3}{l}{\textit{Conventional machine-learning models}} \\
\addlinespace[1pt]

SVM
& $49.67 \pm 1.72$
& $\mathbf{4.43 \pm 0.26}$ \\

KNN
& $15.23 \pm 1.58$
& $\mathbf{0.74 \pm 0.08}$ \\

\addlinespace[2pt]
\midrule
\multicolumn{3}{l}{\textit{Deep-learning models}} \\
\addlinespace[1pt]

ViT-HGR
& $2.61 \pm 0.20$
& $\mathbf{2.50 \pm 0.04}$ \\

WaveFormer
& $6.34 \pm 0.09$
& $\mathbf{5.53 \pm 0.12}$ \\

STCNet
& $4.93 \pm 0.32$
& $\mathbf{4.84 \pm 0.06}$ \\

\bottomrule
\end{tabular}

\vspace{3pt}
\begin{minipage}{0.96\columnwidth}
\footnotesize
\textit{Note:}
Inference time denotes the average downstream-classifier inference time per
sample. Lower values indicate greater inference efficiency. The lower
inference time for each model is shown in bold.
\end{minipage}

\end{table}

\section{Discussion}\label{sec4}
This study presents a fundamentally new perspective on the analysis of sEMG signals by recasting continuous biosignals into discrete, interpretable tokens. Traditional sEMG analysis methods predominantly focus on modeling raw signal waveforms, which often suffer from limited robustness and poor generalizability due to inter-subject variability. In contrast, our tokenization framework prioritizes the extraction of interpretable‌ muscle states, yielding representations that are more consistent with underlying muscle states.

\subsection{Summary of Experimental Results}
The proposed sEMG tokens exhibit three key advantages: inter-subject consistency, interpretability and discriminative representation capacity. 
\textbf{(1)} The proposed tokenization framework demonstrated substantial consistency, with tokens obtained from independently constructed vocabularies achieving an overall Cohen’s Kappa of 0.82. Comparable consistency was also observed at the subject, muscle, and action levels, indicating that the learned token was not dominated by particular individuals, muscles, or actions, but captured muscle-state patterns that were broadly reproducible across different recording conditions. It should be emphasized that tokenization does not completely eliminate inter-subject variability in sEMG signals. Rather, by constructing and sharing a unified vocabulary of muscle states, it maps continuous sEMG signals from different individuals into a common representation space, thereby capturing reproducible patterns of muscle activity.
\textbf{(2)} The proposed sEMG tokens exhibited an interpretable structure related to muscle states. The token feature profiles suggested an organization from high-intensity activation to lower-activation or near-rest states, indicating that the tokens were not merely arbitrary cluster labels. This interpretation was further supported by the occupancy analysis of functional muscle groups: agonist muscles showed higher occupancy of tokens associated with stronger activation states, whereas non-agonist muscles showed higher occupancy of lower-activation or near-rest tokens. Although this does not provide direct physiological validation of each individual token, it provides preliminary evidence that the token representation captures meaningful differences in muscle activation. 
\textbf{(3)} The proposed tokens retained sufficient representational capacity. Across all five classifiers, token-based inputs consistently improved action-recognition performance compared with continuous sEMG signals, indicating that the benefit of tokenization was not limited to a specific model architecture. By mapping local sEMG windows with similar descriptor profiles to shared muscle-state units, the proposed tokenization framework helps reduce the influence of fine-grained waveform fluctuations and subject-specific signal variations while retaining the temporal organization of task-relevant muscle states.

\subsection{Application Potential Enabled by Interpretability}
\begin{figure}
    \centering
    \includegraphics[width=0.9\linewidth]{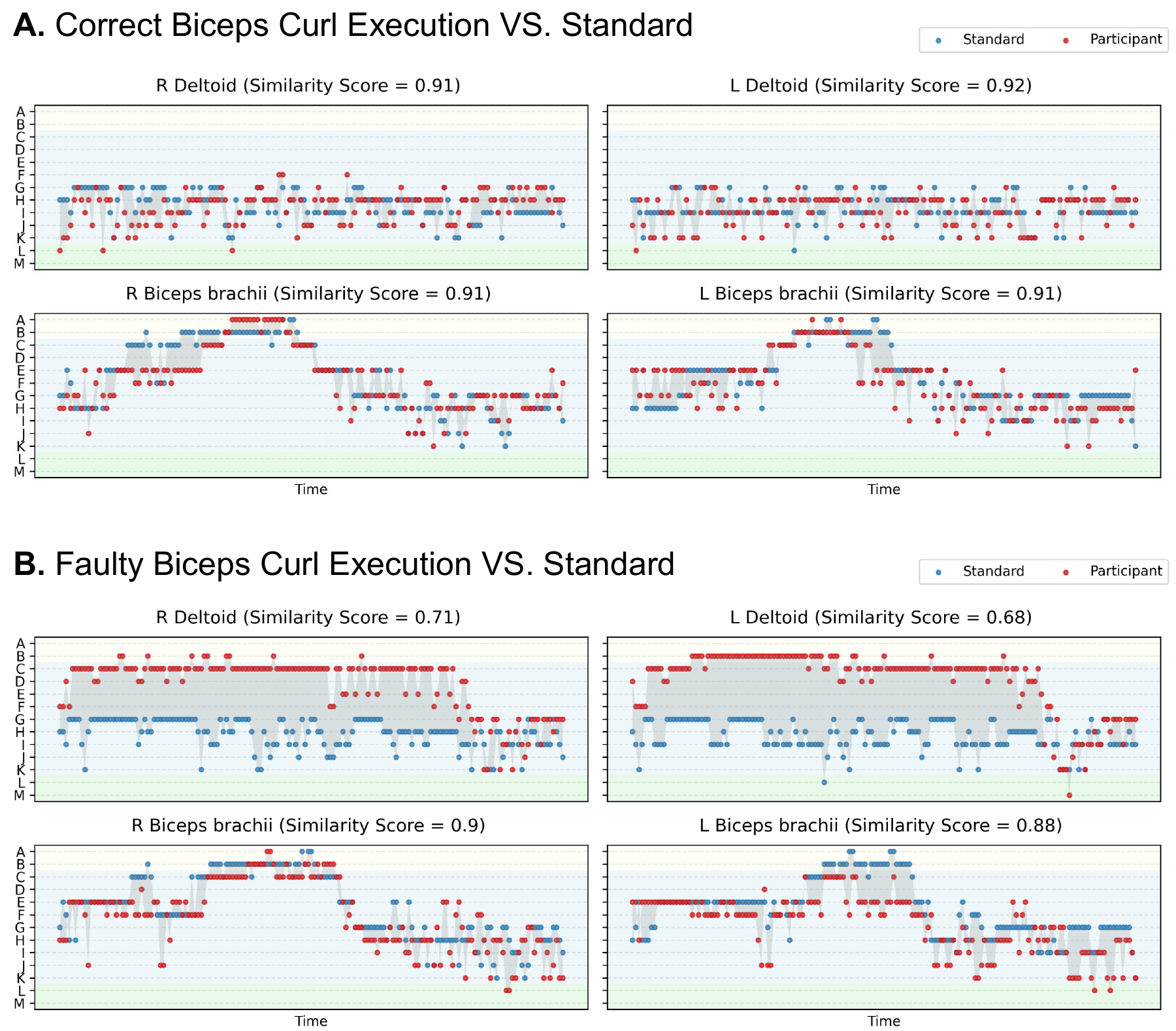}
    \caption{Comparison on sEMG tokens between participant's Biceps Curl with correct execution. Blue points represent the sEMG tokens sequence of the correct execution provided by the fitness expert, while red points indicate the sEMG tokens sequence of an ordinary participant. The shaded regions highlight the discrepancies between the two token sequences, revealing differences in muscle activation states. For visual clarity, this figure focuses on the deltoid and biceps brachii muscles. Full comparisons across all muscle groups are provided in Appendix Fig.~\ref{casestudyall}.}
    \label{casestudy}
\end{figure}

Section \ref{Interpretability} suggests that the token may exhibits an interpretable organization, with different tokens appear to associate with a continuum of muscle activation levels ranging from high-intensity activation to near-rest states. To validate the practical utility of this interpretability, we apply sEMG tokens to movement quality assessment, a task that demands fine-grained insight into execution differences within the same action category. To this end, we conducted a preliminary case study involving expert and novice participants performing correct and intentionally faulty executions of three representative actions. Dynamic Time Warping (DTW) was used to compare the resulting token sequences, both among correct executions and between faulty and correct samples. As shown in Table~\ref{tab:dtw_combined}, both intra-subject and inter-subject analyses consistently show that correct executions generally exhibit higher similarity to expert demonstrations, whereas faulty executions show lower alignment. The clear drop in DTW similarity, captured by this simple metric, shows that the tokens effectively represent subtle differences in muscle activation. The qualitative cases provide a more intuitive interpretation of this result. Figure~\ref{casestudy}-A illustrates the sEMG tokens of an ordinary subject when correctly performing the Biceps Curl. The participant’s sEMG tokens sequence closely aligns with that of the expert, particularly highlighting strong activation in the biceps brachii. In contrast, during the faulty execution in Figure~\ref{casestudy}-B, which involves elbow elevation and compensatory activation, the token sequence exhibits a noticeable increase in deltoid activity, revealing the presence of undesired muscle compensation. This example illustrates how the tokenized representation alone can reveal subtle but physiologically meaningful deviations in muscle recruitment.

These findings suggest the potential rehabilitation-related value of sEMG tokens. By transforming complex continuous sEMG signals into discrete and interpretable state sequences, sEMG tokens may provide clinicians with supplementary information for identifying abnormal recruitment or compensatory muscle activation patterns, thereby supporting movement-quality analysis and rehabilitation assessment. In addition, a shared muscle-state vocabulary maps sEMG signals from different individuals into a common representation space, enabling structured comparisons between individual-specific activation patterns and healthy reference patterns. This comparability may support the quantification of movement deviations and the longitudinal tracking of changes during rehabilitation.

\begin{table*}[t]
\centering
\caption{DTW similarity between sEMG token sequences across intra-subject and cross-subject settings. Values are reported as mean $\pm$ standard deviation.}
\label{tab:dtw_combined}

\setlength{\tabcolsep}{9pt}
\renewcommand{\arraystretch}{1.15}
\small

\begin{tabular}{llcc}
\toprule
\textbf{Setting} & \textbf{Comparison} & \textbf{All Muscles} & \textbf{Primary Muscles} \\
\midrule

\multicolumn{4}{l}{\textbf{(A) Intra-subject: Expert executions}} \\
\midrule

\multirow{2}{*}{High Knees}
& Correct vs.\ Correct
& $0.900 \pm 0.000$
& $0.900 \pm 0.000$ \\
& Faulty vs.\ Correct
& $0.865 \pm 0.000$
& $0.870 \pm 0.006$ \\

\addlinespace[3pt]

\multirow{2}{*}{Squats}
& Correct vs.\ Correct
& $0.907 \pm 0.002$
& $0.942 \pm 0.003$ \\
& Faulty vs.\ Correct
& $0.883 \pm 0.002$
& $0.903 \pm 0.005$ \\

\addlinespace[3pt]

\multirow{2}{*}{Biceps Curl}
& Correct vs.\ Correct
& $0.914 \pm 0.001$
& $0.931 \pm 0.002$ \\
& Faulty vs.\ Correct
& $0.895 \pm 0.001$
& $0.859 \pm 0.002$ \\

\midrule
\multicolumn{4}{l}{\textbf{(B) Cross-subject: Ordinary participants vs Expert reference}} \\
\midrule

\multirow{2}{*}{High Knees}
& Correct vs.\ Expert
& $0.835 \pm 0.003$
& $0.863 \pm 0.004$ \\
& Faulty vs.\ Expert
& $0.820 \pm 0.001$
& $0.793 \pm 0.003$ \\

\addlinespace[3pt]

\multirow{2}{*}{Squats}
& Correct vs.\ Expert
& $0.855 \pm 0.003$
& $0.905 \pm 0.002$ \\
& Faulty vs.\ Expert
& $0.841 \pm 0.004$
& $0.883 \pm 0.006$ \\

\addlinespace[3pt]

\multirow{2}{*}{Biceps Curl}
& Correct vs.\ Expert
& $0.859 \pm 0.001$
& $0.918 \pm 0.001$ \\
& Faulty vs.\ Expert
& $0.816 \pm 0.001$
& $0.770 \pm 0.005$ \\

\bottomrule
\end{tabular}

\vspace{2pt}
\begin{minipage}{0.78\textwidth}
\footnotesize
\textit{Note:} Intra-subject results are computed using multiple executions from the certified expert. Cross-subject results compare ordinary participants against expert references. Primary Muscles focuse on muscles most responsible for the target action and most sensitive to execution errors, and they are defined in Table~\ref{MEQAD}.
\end{minipage}
\end{table*}

\subsection{Limitations and Outlooks}

This study has several limitations that should be acknowledged. 

\textbf{First,} the ActionEMG-43 dataset was collected from only 14 healthy adults, resulting in limited variability in physical function. While this controlled setting was sufficient for the initial evaluation of our proposed sEMG token representation, practical or clinical applications typically involve populations with much greater physiological and motor variability, such as older adults or individuals with motor impairments. Therefore, future studies incorporating more diverse participants are necessary to validate the robustness and generalizability of the proposed method across broader physical conditions. 


\textbf{Second}, the present study evaluated inter-subject variability under a standardized acquisition protocol. All participants performed the same actions, electrodes were placed at fixed locations over the muscle belly of the target muscles, and sufficient rest intervals were introduced between trials to reduce fatigue-related effects. This design reduced acquisition-related variability and enabled a focused evaluation of the proposed sEMG token across individuals under comparable recording conditions. However, practical applications may involve additional sources of variability, including electrode displacement, fatigue accumulation, sensor type, sampling rate, channel configuration, filtering settings, and recording density. For example, HD-sEMG or wrist-based sEMG signals may exhibit signal distributions that differ substantially from the sparse multi-muscle recordings used in this study for whole-body action analysis. These differences may introduce domain shifts and affect token assignments. Future work should further examine the robustness and transferability of the proposed representation across different acquisition conditions and sensor configurations.




Beyond addressing the limitations discussed above, future work could extend the current framework in two directions: representation learning and temporal modeling. First, self-supervised learning could be used to derive latent representations relevant to token generation directly from raw sEMG signals, thereby reducing reliance on handcrafted features and preserving a broader range of information in the resulting token representation. Second, temporal modeling could be incorporated into token sequences to characterize the dynamic evolution of muscle states.

\section{Conclusion}\label{sec5}

This work introduces a novel tokenization framework for sEMG signals, providing a structured and interpretable representation of muscle activation states. In contrast to traditional methods that focus on reconstructing continuous biosignals, our approach emphasizes the extraction of semantically meaningful and compact representations, which are more robust to inter-subject variability and more suitable for downstream applications. Through extensive experiments on consistency analysis, interpretability evaluation, and human action recognition, we demonstrate that sEMG tokens exhibit strong inter-subject consistency, interpretability, and representation capacity. Overall, this study presents a robust and interpretable sEMG tokenization framework, highlighting the potential of discrete muscle-state representations for building explainable muscle–computer interfaces in rehabilitation, personalized training, and human–machine interaction.





\printcredits

\section*{Declaration of competing interest}
The authors declare no competing interests.
The authors declare that they have no known competing financial interests or personal relationships that could have appeared to influence the work reported in this paper.

\section*{Data availability}
The ActionEMG-43 dataset used in this study is currently available from the corresponding author on reasonable request. The dataset will be released publicly after the manuscript is accepted for publication, to support reproducibility and further research.

\section*{Acknowledgments}
This work was supported by the Tsinghua University Initiative Scientific Research Program and the Natural Science Foundation of Beijing(Grant No.L242049.

\section*{Ethics}
This study was approved by the Institutional Review Board of Tsinghua University (Approval No.  THU01-20240069).  All the data was totally de-identiﬁed to ensure con-ﬁdentiality and privacy.

\appendix
\section{Appendix}
\subsection{Dataset Description}

The ActionEMG-43 dataset contains 43 distinct human actions commonly involved in physical training, rehabilitation, and daily movement tasks. A detailed list of all action classes is presented in Table.~\ref{ActionEMG-43}. In addition to general actions, we constructed an Movement Quality Evaluation Dataset consisting of representative exercises, each with a well-defined standard execution template and a common error variant. To facilitate objective evaluation, we analyzed muscle activation differences between correct and incorrect executions. Table.~\ref{MEQAD} summarizes the key characteristics of three representative actions, including definitions, error descriptions, and associated changes in muscle activity.

\begin{table*}[t]
\centering
\caption{Action Categories in the ActionEMG-43 Dataset}
\label{ActionEMG-43}

\setlength{\tabcolsep}{6pt}
\renewcommand{\arraystretch}{1.15}
\small

\begin{tabular}{cc|cc}
\toprule
\textbf{ID} & \textbf{Action Name} & \textbf{ID} & \textbf{Action Name} \\
\midrule

0  & Jogging  & 22 & Plank with Alternating Arm Raise \\
1  & High Knees  & 23 & Standing Head Turn to Look at Object \\
2  & Frog Jumps  & 24 & Maximal Forward Reach with One Arm Fist Clench \\
3  & High Kicks  & 25 & Lunge Jumps \\
4  & Squats  & 26 & Single-Leg Stand \\
5  & Walking with Turning  & 27 & Biceps Curl \\
6  & Sit Up  & 28 & Horse Stance \\
7  & Butt Kicks  & 29 & Alternating Box Jumps \\
8  & Side Kick  & 30 & V-Up \\
9  & Fast Feet with Arm Swing and Squat  & 31 & Straight Leg Lunge \\
10 & Foot Fire  & 32 & Bounding \\
11 & Lateral Lunge  & 33 & Arm Circles \\
12 & Quadruped Bridge  & 34 & Lat Pulldown without Equipment \\
13 & Hurdle Steps  & 35 & Brisk Walking \\
14 & Side Leg Swings  & 36 & Forward Bend to Pick Up Object \\
15 & Jumping Jacks  & 37 & Heel-to-Glute with Arm Reach \\
16 & Seated Toe Raises  & 38 & Standing Overhead Press without Equipment \\
17 & Walking Lunge with Trunk Rotation  & 39 & Cross Steps \\
18 & Abdominal and Back Stretch  & 40 & Calf Raises \\
19 & Standing Side Bend  & 41 & Standing Front Press without Equipment \\
20 & Bent Over T-Raises  & 42 & Standing External Hip Rotation \\
21 & Standing Head-Hand Resistance Stretch &  &  \\
\bottomrule
\end{tabular}
\end{table*}

\begin{table*}[t]
\centering
\caption{Movement Quality Assessment Dataset (MEQAD)}
\label{MEQAD}

\setlength{\tabcolsep}{6pt}
\renewcommand{\arraystretch}{1.2}
\small

\begin{tabular}{p{1.2cm} p{4.2cm} p{4.2cm} p{4.2cm}}
\toprule
\textbf{Action Name} 
& \textbf{Correct Action Definition} 
& \textbf{Faulty Action Description} 
& \textbf{Primary Muscle Activation Differences} \\
\midrule

High Knees  
& Performed with \textbf{forefoot landing}, an upright and stable torso, and consistent rhythmic stepping throughout the movement.  
& Whole-foot contact occurs at landing, indicating reduced lower-limb control and insufficient propulsion.  
& Reduced activation of the \textbf{gastrocnemius} due to the transition from forefoot to flat-foot landing.  
\\

Squats  
& Moderate forward trunk inclination with coordinated hip and knee flexion, heels firmly grounded, and posteriorly maintained center of gravity.  
& Limited ankle dorsiflexion causes heel lift, shifting the center of mass anteriorly and reducing postural stability.  
& Compensatory overactivation of the \textbf{gastrocnemius}, reflecting impaired ankle mobility control.  
\\

Biceps Curl  
& Elbows remain close to the torso, with elbow flexion driven primarily by the \textbf{biceps brachii}.  
& Excessive shoulder flexion introduces forward displacement of the upper arm, disrupting the intended motion trajectory.  
& Increased activation of the \textbf{anterior deltoid}, indicating compensatory recruitment of non-target muscles.  
\\

\bottomrule
\end{tabular}
\end{table*}

\subsection{Definitions of sEMG Features}

The Table~\ref{tab:semg_feature_definitions}  summarizes the mathematical definitions and signal-level interpretations of all sEMG features used in the proposed framework.

\begin{table*}[t]
\centering
\caption{
Mathematical definitions and description of the selected
sEMG features. For a windowed sEMG segment,
$x_w$ denotes the $w$-th sample and $W$ is the number of samples.
For the frequency-domain features, $P_j$ denotes the estimated power
spectral density at frequency bin $f_j$, and $M$ is the number of
frequency bins considered.
}
\label{tab:semg_feature_definitions}

\renewcommand{\arraystretch}{1.05}
\setlength{\tabcolsep}{4pt}
\scriptsize

\begin{tabular}{
    >{\centering\arraybackslash}m{3.0cm}
    >{\centering\arraybackslash}m{4.90cm}
    >{\raggedright\arraybackslash}m{7.00cm}
}
\toprule
\textbf{Feature}
&
\textbf{Mathematical definition}
&
\textbf{Description}
\\
\midrule

Root Mean Square
&
$\displaystyle
\mathrm{RMS}
=
\sqrt{
\frac{1}{W}
\sum_{w=1}^{W}x_w^2
}
$
&
RMS is a popular feature in analysis of the EMG signal. It stands to the square root of the mean square of the signal amplitude.
\\
\cmidrule(lr){1-3}

Mean Absolute Value
&
$\displaystyle
\mathrm{MAV}
=
\frac{1}{W}
\sum_{w=1}^{W}|x_w|
$
&
MAV is one of the most popular used in EMG signal analysis. It is an average of absolute value of the sEMG signal amplitude in a segment.
\\
\cmidrule(lr){1-3}

Waveform Length
&
$\displaystyle
\mathrm{WL}
=
\sum_{w=1}^{W-1}
|x_{w+1}-x_w|
$
&
WL is a measure of complexity of the sEMG signal. It is defined as cumulative length of the EMG waveform over the time segment.
\\
\cmidrule(lr){1-3}

Zero Crossing
&
$\displaystyle
\begin{aligned}
\mathrm{ZC}
&=
\left\{
\begin{array}{l}
\displaystyle
\sum_{w=1}^{W-1}
\operatorname{sgn}
\left(
\mathbf{x}_{w}\cdot\mathbf{x}_{w-1}
\right)
\\[3pt]
\left(
\mathbf{x}_{w}-\mathbf{x}_{w-1}
\right)
\geq\epsilon
\end{array}
\right.
\\[3pt]
\text{and} 
\operatorname{sgn}(x)
&=
\left\{
\begin{array}{ll}
1, & \text{if }x<\epsilon,\\
0, & \text{otherwise}.
\end{array}
\right.
\end{aligned}
$
&
ZC is a measure of frequency information of the sEMG signal that is defined in time domain. It is a number of times that amplitude values of the EMG signal cross zero amplitude level.
\\
\cmidrule(lr){1-3}

Slope Sign Change
&
$\displaystyle
\begin{aligned}
\mathrm{SSC}
={}&
\sum_{w=2}^{W-1}
f\left[
\left(
\mathbf{x}_{w}-\mathbf{x}_{w-1}
\right)
\cdot
\left(
\mathbf{x}_{w}-\mathbf{x}_{w+1}
\right)
\right],
\\[3pt]
\text{and}
f(x)
&=
\left\{
\begin{array}{ll}
1, & \text{if }x\geq\epsilon,\\
0, & \text{otherwise}.
\end{array}
\right.
\end{aligned}
$
&
SSC is another method to represent frequency information of the sEMG signal. It is a number of times that slope of the sEMG signal changes sign. The number of changes between the positive and negative slopes
among three sequential segments is performed with the threshold function for avoiding background noise in the sEMG signal.
\\
\cmidrule(lr){1-3}

Willison Amplitude
&
$\displaystyle
\begin{aligned}
\mathrm{WAMP}
={}&
\sum_{w=1}^{W-1}
f\left(
|\mathbf{x}_{w}-\mathbf{x}_{w-1}|
\right),
\\[3pt]
\text{and}
f(x)
&=
\left\{
\begin{array}{ll}
1, & \text{if }x\geq\epsilon,\\
0, & \text{otherwise}.
\end{array}
\right.
\end{aligned}
$
&
WAMP measures frequency-related information in the sEMG signal, similar to the zero-crossing feature. It counts the number of times the amplitude difference between two adjacent sEMG samples exceeds a predefined threshold. Moreover, WAMP is associated with motor-unit action potential activity and muscle contraction force.
\\
\cmidrule(lr){1-3}

Mean Frequency
&
$\displaystyle
\mathrm{MNF}
=
\frac{
\sum_{j=1}^{M} f_j P_j
}{
\sum_{j=1}^{M} P_j
}
$
&
MNF is the average frequency of the sEMG power spectrum.
It is calculated as the sum of the products of the spectral power and
the corresponding frequency, divided by the total spectral power. MNF is also referred to as the central frequency or the spectral center of gravity, where $f_j$ denotes the frequency at the $j$-th frequency bin, $P_j$ denotes the EMG power spectrum at the $j$-th frequency bin, and $M$ denotes the total number of frequency bins.
\\
\cmidrule(lr){1-3}

Median Frequency
&
$\displaystyle
\sum_{j=1}^{\mathrm{MDF}} P_j
=
\sum_{j=\mathrm{MDF}}^{M} P_j
=
\frac{1}{2}
\sum_{j=1}^{M} P_j
$
&
MDF is a frequency at which the spectrum is divided into two regions with equal amplitude.
\\
\cmidrule(lr){1-3}

Power Spectrum Ratio
&
$\displaystyle
\mathrm{PSR}
=
\frac{P_0}{P}
=
\frac{
\sum_{j=f_0-n}^{f_0+n} P_j
}{
\sum_{j=-\infty}^{\infty} P_j
}
$
&
The PSR is defined as the ratio between the energy
$P_0$ in the neighborhood of the maximum value of the sEMG power spectrum
and the total energy $P$ of the entire EMG power spectrum
\\
\cmidrule(lr){1-3}

Auto-Regressive Coefficients
&
$\displaystyle
x_i
=
\sum_{p=1}^{P} a_p\,x_{i-p}
+
w_i
$
&
The autoregressive (AR) model is a prediction model that represents each
sample of the EMG signal, $x_i$, as a linear combination of the previous
samples, $x_{i-p}$, together with a white-noise error term $w_i$. $P$ is the order of the AR model.
\\

\bottomrule
\end{tabular}
\end{table*}
\subsection{Detailed Results of the Sensitivity Analysis on the Number of Muscle-State Units}
We report the complete results for all values of $K \in [2,26]$, including Cohen’s Kappa, Accuracy, Macro-P, Macro-R, and Macro-F1 for both SVM and ViT-HGR across five cross-subject folds. The full table is provided in Table~\ref{tab:k_sensitivity}.

\begin{table*}[t]
\centering
\caption{Effect of the number of muscle-state units $K$ on token consistency and classification performance. Results are reported as mean $\pm$ standard deviation over five folds. Bold values indicate the highest mean in each metric.}
\label{tab:k_sensitivity}

\setlength{\tabcolsep}{3.0pt}
\renewcommand{\arraystretch}{1.00}
\scriptsize

\resizebox{\textwidth}{!}{%
\begin{tabular}{c c cccc cccc}
\toprule
\multirow{2}{*}{$K$}
& \multirow{2}{*}{Cohen's $\kappa$}
& \multicolumn{4}{c}{SVM}
& \multicolumn{4}{c}{ViT-HGR} \\
\cmidrule(lr){3-6}
\cmidrule(lr){7-10}
&
&
Accuracy
& Macro-P
& Macro-R
& Macro-F1
& Accuracy
& Macro-P
& Macro-R
& Macro-F1 \\
\midrule

2
& $0.94 \pm 0.03$
& $0.3458 \pm 0.0693$
& $0.3521 \pm 0.0711$
& $0.3148 \pm 0.0857$
& $0.2966 \pm 0.0732$
& $0.4105 \pm 0.0518$
& $0.4160 \pm 0.0473$
& $0.3824 \pm 0.0587$
& $0.3619 \pm 0.0537$ \\

3
& $\mathbf{0.95 \pm 0.03}$
& $0.5693 \pm 0.0392$
& $0.5795 \pm 0.0491$
& $0.5391 \pm 0.0318$
& $0.5244 \pm 0.0408$
& $0.6519 \pm 0.0541$
& $0.6501 \pm 0.0564$
& $0.6321 \pm 0.0544$
& $0.6098 \pm 0.0570$ \\

4
& $0.58 \pm 0.30$
& $0.6287 \pm 0.0513$
& $0.6629 \pm 0.0552$
& $0.6160 \pm 0.0506$
& $0.6040 \pm 0.0474$
& $0.7053 \pm 0.0505$
& $0.7187 \pm 0.0419$
& $0.6917 \pm 0.0435$
& $0.6737 \pm 0.0446$ \\

5
& $0.93 \pm 0.03$
& $0.6397 \pm 0.0567$
& $0.6636 \pm 0.0524$
& $0.6336 \pm 0.0524$
& $0.6184 \pm 0.0574$
& $0.7156 \pm 0.0477$
& $0.7262 \pm 0.0426$
& $0.7015 \pm 0.0447$
& $0.6842 \pm 0.0449$ \\

6
& $0.69 \pm 0.26$
& $0.6421 \pm 0.0504$
& $0.6606 \pm 0.0425$
& $0.6260 \pm 0.0426$
& $0.6082 \pm 0.0445$
& $0.7170 \pm 0.0466$
& $0.7419 \pm 0.0579$
& $0.7048 \pm 0.0436$
& $0.6890 \pm 0.0500$ \\

7
& $0.84 \pm 0.04$
& $0.6490 \pm 0.0380$
& $0.6834 \pm 0.0428$
& $0.6504 \pm 0.0334$
& $0.6478 \pm 0.0399$
& $0.7324 \pm 0.0469$
& $0.7502 \pm 0.0509$
& $0.7160 \pm 0.0470$
& $0.7025 \pm 0.0486$ \\

8
& $0.84 \pm 0.02$
& $0.6494 \pm 0.0459$
& $0.6881 \pm 0.0436$
& $0.6488 \pm 0.0348$
& $0.6499 \pm 0.0394$
& $0.7418 \pm 0.0325$
& $0.7614 \pm 0.0278$
& $0.7265 \pm 0.0258$
& $0.7149 \pm 0.0276$ \\

9
& $0.85 \pm 0.02$
& $0.6570 \pm 0.0375$
& $0.6866 \pm 0.0384$
& $0.6666 \pm 0.0388$
& $0.6686 \pm 0.0346$
& $0.7423 \pm 0.0238$
& $0.7599 \pm 0.0236$
& $0.7276 \pm 0.0193$
& $0.7132 \pm 0.0236$ \\

10
& $0.82 \pm 0.06$
& $0.6641 \pm 0.0314$
& $0.6968 \pm 0.0403$
& $0.6739 \pm 0.0356$
& $0.6704 \pm 0.0345$
& $0.7409 \pm 0.0278$
& $0.7499 \pm 0.0246$
& $0.7279 \pm 0.0268$
& $0.7134 \pm 0.0273$ \\

11
& $0.75 \pm 0.11$
& $\mathbf{0.6954 \pm 0.0380}$
& $\mathbf{0.7258 \pm 0.0348}$
& $\mathbf{0.6961 \pm 0.0391}$
& $\mathbf{0.6820 \pm 0.0426}$
& $0.7505 \pm 0.0256$
& $0.7671 \pm 0.0335$
& $0.7362 \pm 0.0215$
& $0.7233 \pm 0.0248$ \\

12
& $0.80 \pm 0.10$
& $0.6881 \pm 0.0390$
& $0.7189 \pm 0.0396$
& $0.6872 \pm 0.0379$
& $0.6732 \pm 0.0357$
& $0.7501 \pm 0.0310$
& $0.7633 \pm 0.0363$
& $0.7329 \pm 0.0284$
& $0.7185 \pm 0.0323$ \\

13
& $0.82 \pm 0.09$
& $0.6790 \pm 0.0339$
& $0.7188 \pm 0.0394$
& $0.6933 \pm 0.0311$
& $0.6782 \pm 0.0323$
& $0.7550 \pm 0.0248$
& $0.7620 \pm 0.0309$
& $0.7386 \pm 0.0227$
& $0.7244 \pm 0.0248$ \\

14
& $0.80 \pm 0.09$
& $0.6798 \pm 0.0328$
& $0.7174 \pm 0.0464$
& $0.6803 \pm 0.0369$
& $0.6660 \pm 0.0386$
& $0.7592 \pm 0.0257$
& $0.7667 \pm 0.0323$
& $0.7446 \pm 0.0256$
& $0.7294 \pm 0.0271$ \\

15
& $0.79 \pm 0.10$
& $0.6856 \pm 0.0363$
& $0.7088 \pm 0.0331$
& $0.6854 \pm 0.0364$
& $0.6708 \pm 0.0350$
& $0.7525 \pm 0.0237$
& $0.7733 \pm 0.0336$
& $0.7364 \pm 0.0238$
& $0.7278 \pm 0.0270$ \\

16
& $0.72 \pm 0.11$
& $0.6856 \pm 0.0368$
& $0.7150 \pm 0.0240$
& $0.6873 \pm 0.0296$
& $0.6728 \pm 0.0274$
& $\mathbf{0.7619 \pm 0.0224}$
& $0.7770 \pm 0.0276$
& $\mathbf{0.7463 \pm 0.0198}$
& $\mathbf{0.7365 \pm 0.0208}$ \\

17
& $0.77 \pm 0.10$
& $0.6809 \pm 0.0432$
& $0.7121 \pm 0.0261$
& $0.6887 \pm 0.0350$
& $0.6738 \pm 0.0323$
& $0.7550 \pm 0.0187$
& $0.7780 \pm 0.0233$
& $0.7410 \pm 0.0165$
& $0.7303 \pm 0.0177$ \\

18
& $0.73 \pm 0.12$
& $0.6826 \pm 0.0430$
& $0.7111 \pm 0.0324$
& $0.6891 \pm 0.0407$
& $0.6767 \pm 0.0379$
& $0.7571 \pm 0.0140$
& $\mathbf{0.7795 \pm 0.0230}$
& $0.7414 \pm 0.0102$
& $0.7312 \pm 0.0115$ \\

19
& $0.77 \pm 0.06$
& $0.6820 \pm 0.0489$
& $0.7132 \pm 0.0309$
& $0.6828 \pm 0.0336$
& $0.6690 \pm 0.0313$
& $0.7554 \pm 0.0183$
& $0.7697 \pm 0.0250$
& $0.7387 \pm 0.0181$
& $0.7280 \pm 0.0214$ \\

20
& $0.72 \pm 0.11$
& $0.6820 \pm 0.0412$
& $0.7239 \pm 0.0412$
& $0.6818 \pm 0.0391$
& $0.6695 \pm 0.0367$
& $0.7567 \pm 0.0204$
& $0.7666 \pm 0.0197$
& $0.7407 \pm 0.0189$
& $0.7285 \pm 0.0199$ \\

21
& $0.76 \pm 0.07$
& $0.6766 \pm 0.0361$
& $0.7120 \pm 0.0296$
& $0.6764 \pm 0.0386$
& $0.6636 \pm 0.0370$
& $0.7504 \pm 0.0231$
& $0.7696 \pm 0.0201$
& $0.7339 \pm 0.0215$
& $0.7223 \pm 0.0245$ \\

22
& $0.75 \pm 0.11$
& $0.6772 \pm 0.0357$
& $0.7106 \pm 0.0254$
& $0.6757 \pm 0.0309$
& $0.6644 \pm 0.0327$
& $0.7525 \pm 0.0213$
& $0.7758 \pm 0.0186$
& $0.7381 \pm 0.0193$
& $0.7257 \pm 0.0219$ \\

23
& $0.75 \pm 0.14$
& $0.6811 \pm 0.0363$
& $0.7202 \pm 0.0399$
& $0.6791 \pm 0.0403$
& $0.6691 \pm 0.0419$
& $0.7579 \pm 0.0231$
& $0.7686 \pm 0.0267$
& $0.7417 \pm 0.0222$
& $0.7289 \pm 0.0213$ \\

24
& $0.61 \pm 0.11$
& $0.6704 \pm 0.0369$
& $0.7162 \pm 0.0286$
& $0.6729 \pm 0.0356$
& $0.6587 \pm 0.0332$
& $0.7581 \pm 0.0234$
& $0.7715 \pm 0.0379$
& $0.7418 \pm 0.0254$
& $0.7335 \pm 0.0271$ \\

25
& $0.55 \pm 0.04$
& $0.6802 \pm 0.0356$
& $0.7139 \pm 0.0270$
& $0.6778 \pm 0.0317$
& $0.6665 \pm 0.0321$
& $0.7538 \pm 0.0193$
& $0.7633 \pm 0.0285$
& $0.7371 \pm 0.0188$
& $0.7254 \pm 0.0223$ \\

\bottomrule
\end{tabular}%
}

\end{table*}

\subsection{Case Study: sEMG Tokens Analysis of Biceps Curl Across All Muscles}

\begin{figure}
    \includegraphics[width=1\linewidth]{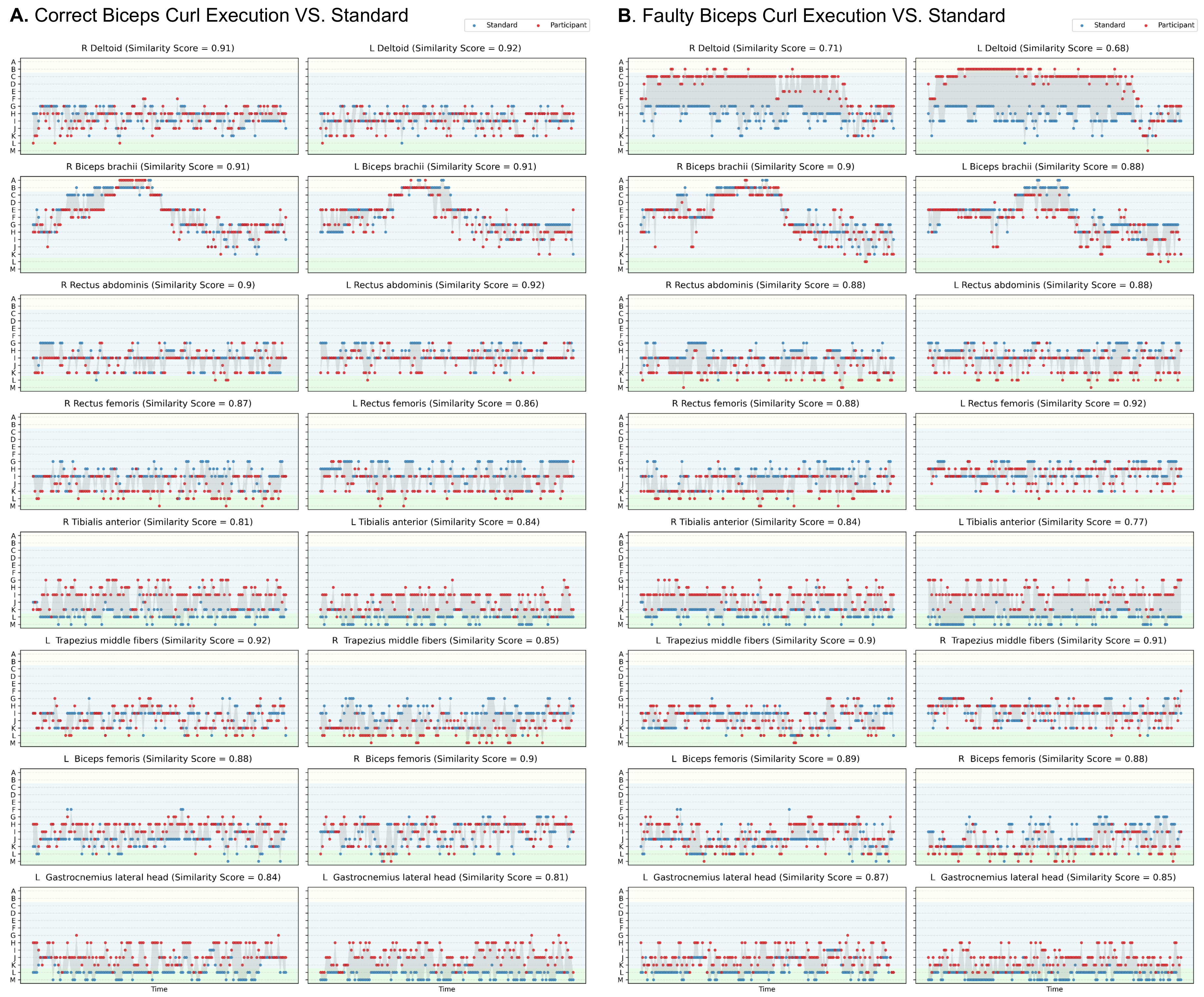}
    \caption{sEMG Tokens-Based Comparison between Ordinary Individuals' Biceps Curl Execution and Expert-Level Performance. Blue points represent the sEMG tokens transitions of the fitness expert during a single execution, while red points indicate the sEMG tokens transitions of the ordinary participant during a single execution. The shaded regions highlight the discrepancies between the two token sequences, revealing differences in muscle activation states.}
    \label{casestudyall}
\end{figure}

To extend the qualitative findings in the main text, this appendix presents additional sEMG tokens comparisons across all recorded muscle groups. Figures~\ref{casestudyall} illustrateS the token trajectories of both the fitness expert and the ordinary participant during a single biceps curl execution. These visualizations reveal muscle-specific deviations, such as compensatory activation in non-target muscles like the deltoid or insufficient recruitment of primary movers, which are often difficult to detect in raw sEMG signals. By capturing differences in muscle states, the tokenized representation provides a structured and interpretable perspective on inter-muscle coordination, enabling a more comprehensive assessment of execution quality.


\bibliographystyle{cas-model2-names}

\bibliography{cas-refs}



\end{document}